\documentclass[journal]{vgtc}                % final (journal style)
\ifpdf%                                % if we use pdflatex
  \pdfoutput=1\relax                   % create PDFs from pdfLaTeX
  \usepackage{graphicx}                % allow us to embed graphics files
  \DeclareGraphicsExtensions{.pdf,.png,.jpg,.jpeg} % for pdflatex we expect .pdf, .png, or .jpg files
\else%                                 % else we use pure latex
  \usepackage{graphicx}                % allow us to embed graphics files
  \DeclareGraphicsExtensions{.eps}     % for pure latex we expect eps files
\fi%

\graphicspath{{figures/}{pictures/}{images/}{./}} % where to search for the images

\usepackage{microtype}                 % use micro-typography (slightly more compact, better to read)
\PassOptionsToPackage{warn}{textcomp}  % to address font issues with \textrightarrow
\usepackage{textcomp}                  % use better special symbols
\usepackage{mathptmx}                  % use matching math font
\usepackage{times}                     % we use Times as the main font
\usepackage{cite}                      % needed to automatically sort the references
\usepackage{tabu}                      % only used for the table example
\usepackage{booktabs}                  % only used for the table example

\usepackage[table,xcdraw]{xcolor}
\usepackage{xcolor}
\usepackage{multirow}
\usepackage{subcaption}
\newcommand{\revised}[1]{{\color{black}{#1}}}

\newcommand{\leilani}[1]{{\color{violet}{\bf (Leilani: #1)}}}

\usepackage{fontawesome}
\usepackage[shortlabels]{enumitem}
\usepackage{makecell}
\usepackage{graphicx}
\usepackage{times}
\usepackage{comment}
\usepackage{amsmath}
\usepackage{array}
\newcolumntype{L}[1]{>{\raggedright\let\newline\\\arraybackslash\hspace{0pt}}m{#1}}
\newcolumntype{C}[1]{>{\centering\let\newline\\\arraybackslash\hspace{0pt}}m{#1}}
\newcolumntype{R}[1]{>{\raggedleft\let\newline\\\arraybackslash\hspace{0pt}}m{#1}}
\definecolor{bird}{HTML}{F8766C}
\definecolor{movie}{HTML}{15BFC4}
\newcommand\crule[3][black]{\textcolor{#1}{\rule{#2}{#3}}}

\usepackage[bookmarks,backref=true,linkcolor=black]{hyperref} %,colorlinks
\hypersetup{
  pdfauthor = {},
  pdftitle = {},
  pdfsubject = {},
  pdfkeywords = {},
  colorlinks=true,
  linkcolor= black,
  citecolor= black,
  pageanchor=true,
  urlcolor = black,
  plainpages = false,
  linktocpage
}
\onlineid{1199}

\vgtccategory{Research}
\vgtcpapertype{theory/model}

\title{An Evaluation-Focused Framework for Visualization Recommendation Algorithms}

\author{Zehua Zeng, Phoebe Moh, Fan Du, Jane Hoffswell, Tak Yeon Lee, Sana Malik, Eunyee Koh, and Leilani Battle}

\authorfooter{
\item
 Zehua Zeng is with University of Maryland. E-mail: zhzeng@umd.edu.
\item
 Phoebe Moh is with University of Maryland. E-mail: pmoh@umd.edu.
\item
 Fan Du is with Adobe Research. Email: fdu@adobe.com.
\item
 Jane Hoffswell is with Adobe Research. Email: jhoffs@adobe.com.
\item
 Tak Yeon Lee is with KAIST, work performed while at Adobe Research. Email: takyeonlee@kaist.ac.kr.
\item
 Sana Malik is with Adobe Research. Email: sanmalik@adobe.com.
\item
 Eunyee Koh is with Adobe Research. Email: eunyee@adobe.com.
\item
 Leilani Battle is with University of Washington, work performed while at University of Maryland. E-mail: leibatt@uw.edu.
}

\shortauthortitle{Biv \MakeLowercase{\textit{et al.}}: Global Illumination for Fun and Profit}
\abstract{
Although we have seen a proliferation of algorithms for recommending visualizations, these algorithms are rarely compared with one another, making it difficult to ascertain which algorithm is best for a given visual analysis scenario.
Though several formal frameworks have been proposed in response, we believe this issue persists because visualization recommendation algorithms are inadequately specified from an \emph{evaluation} perspective.
In this paper, we propose an evaluation-focused framework to contextualize and compare a broad range of visualization recommendation algorithms.
We present the structure of our framework, where algorithms are specified using three components: (1)~a graph representing the full space of possible visualization designs, (2)~the method used to traverse the graph for potential candidates for recommendation, and (3)~an oracle used to rank candidate designs.
To demonstrate how our framework guides the formal comparison of algorithmic performance, we not only theoretically compare five existing representative recommendation algorithms, but also empirically compare four new algorithms generated based on our findings from the theoretical comparison.
Our results show that these algorithms behave similarly in terms of user performance, highlighting the need for more rigorous formal comparisons of recommendation algorithms to further clarify their benefits in various analysis scenarios.

} 

\begin{document}


\firstsection{Introduction}

\maketitle

\vspace{-1mm}
The visualization community has developed a wide variety of systems for recommending how to visualize data~\cite{Zhu2020survey}.
The algorithms behind these systems aim to help users uncover meaningful insights in their data by automatically generating visualizations for analysts to explore.
For example, Voyager~\cite{Wongsuphasawat2015voyager, Wongsuphasawat2017voyager2} encourages broad data exploration by recommending effective charts based on Mackinlay's~\cite{Mackinlay1986automating} design principles.
VizDeck~\cite{Key2012vizdeck} and Foresight~\cite{Demiralp2017foresight} recommend visualizations based on standard statistical characteristics of the dataset.
SeeDB~\cite{Vartak2015seedb} recommends visualizations based on a self-defined criterion of statistical ``interestingness'', or divergence of a sub-population from the whole.

While this panoply of recommendation algorithms provides many viable alternatives, it is unclear which algorithm should be prioritized for any given visualization scenario.
%this panoply of algorithms also makes picking the most effective recommendation algorithm difficult for data analysts. 
In a review of existing evaluation practices, we find that many recommendation systems evaluate their recommendation algorithms in isolation~\cite{Luo2018deepeye}, or construct benchmarks that their systems are already optimized for~\cite{Mackinlay1986automating, Vartak2015seedb, Wongsuphasawat2015voyager, Wongsuphasawat2017voyager2}.
Even evaluations that do compare different algorithms do not measure user performance~\cite{Hu2019vizml, Lin2020dziban, Moritz2018formalizing}.
In other words, our community tends to generate new visualization recommendation algorithms without giving commensurate thought on how to evaluate them. As a result, the visualization community lacks rigorous theoretical and empirical guidance for how and when to apply each of these algorithms effectively.

One way to address this \revised{problem} is to develop a standardized framework for comparing different visualization recommendation algorithms.
%With a framework, we could rank the efficacy of existing recommendation algorithms for different visual analysis scenarios, and reason about the theoretical benefits of new algorithms to be developed in the future.
Given that the purpose of these algorithms is to help analysts visually explore their data, a standardized framework should enable us to directly compare algorithms based on how they impact a user's performance for a variety of visual analysis tasks\revised{~\cite{Kim2018assessing, Amar2005lowlevel}}.
The framework should also facilitate comparison of the algorithmic performance of the proposed approaches; for example, the framework should enable us to compare how each algorithm enumerates and traverses the design space of candidate visualizations in search of an optimal recommendation.
%However, there is also a cost to searching the visualization design space; ideally, these algorithms should also minimize how much of the design space they traverse to find optimal recommendations. 

% \noindent\jane{Rewrote the last sentence here to frame it in terms of the overall goals of the framework (our main contribution), rather than as a general goal that recommendation algorithms are attempting to satisfy. Does this seem like a fair rephrasing? I also think the example shared by Eunyee in the comments is worth considering for how we discuss the recommendation approaches more generally.}

In this paper, we propose an evaluation-focused framework to enable more effective theoretical and empirical comparisons of visualization recommendation algorithms.
Our framework is based on the central process connecting most if not all of these algorithms: to generate the ``best'' recommendations, an algorithm must be able to \textbf{enumerate} the space of possible visualization designs and \textbf{rank} this design space, often by approximating and comparing the utility 
% \jane{instead of value, maybe utility?} 
of candidate visualizations.
Our evaluation framework is defined through three major components:
(1) a network representing the space of all possible visualization designs for a given dataset, where nodes are visualization designs and edges connect designs that differ by a single encoding 
% \jane{encoding? attribute? what do you mean when you say a visualization differs by a single design. maybe switching the order to "one data or design transformation"} 
or data transformation; (2) the method a recommendation algorithm uses to traverse the design space to enumerate candidate visualization designs; and (3) an oracle used to approximate and rank the value of candidate visualizations that are enumerated.

%We observe that, although the existing visualization recommendation systems utilize different recommendation algorithms, they follow a similar recommendation process with  three main steps: enumerating, searching, then ranking.
Existing frameworks such as CompassQL~\cite{Wongsuphasawat2016towards}, ZQL~\cite{Siddiqui2017fast}, and Draco~\cite{Moritz2018formalizing}, focus on generating new visualization recommendation algorithms, 
rather than comparing algorithms.
As a result, behavioral differences are not intuitively captured through these frameworks, making it difficult to reason about the differences in algorithmic performance.
For example, it is not clear how one might cluster different recommendation algorithms based on their behavioral similarity.
With our framework, these behavioral differences become obvious.
%Nevertheless, various recommendation systems utilize different searching methods.
For example, Voyager~\cite{Wongsuphasawat2015voyager, Wongsuphasawat2017voyager2} 
% and Dziban~\cite{Lin2020dziban} 
by default recommends visualizations which are one design or data transformation away from the current visualization, representing a narrow but efficient traversal of the  visualization design space. In contrast, machine-learning-based algorithms enumerate and rank massive sub-spaces of visualization designs represented by the model's input features~\cite{Hu2019vizml, Luo2018deepeye}.

We demonstrate the generality and coverage provided by our framework by comparing the behavior of five visualization recommendation systems: Voyager~\cite{Wongsuphasawat2015voyager, Wongsuphasawat2017voyager2}, DeepEye~\cite{Luo2018deepeye}, Foresight~\cite{Demiralp2017foresight}, Show Me~\cite{Mackinlay2007showme} and Dziban~\cite{Lin2020dziban}.
We also show how our framework clarifies gaps in the literature where new algorithms can be formed, simply by varying traversal method and oracle combinations.
Using two common graph traversal methods, breadth-first search (BFS) and depth-first search (DFS), and the oracles for Voyager~\cite{Wongsuphasawat2015voyager}
%CompassQL~\cite{Wongsuphasawat2016towards}
% (the one was used in Voyagers~\cite{Wongsuphasawat2015voyager, Wongsuphasawat2017voyager2}) 
and Dziban~\cite{Lin2020dziban}, we construct four recommendation algorithms: CompassQL+BFS (i.e., Voyager), CompassQL+DFS, Dziban+BFS, and Dziban+DFS.
We then use our framework to design a user study to guide the empirical evaluation of these four visualization recommendation algorithms. 
% \jane{add comment about results}
Our results show that subjects did not perform significantly better with Dziban compared to CompassQL in focused-oriented tasks, however subjects did find Dziban's recommendations to be more intuitive in post-task survey ratings.
These findings reinforce our argument that we need more evaluation-focused frameworks to elucidate the benefits of existing recommendation algorithms in real-world visual analysis scenarios. All of our data and code are available online on OSF: \url{https://osf.io/txqsu/}.
\vspace{-2mm}
\section{Related Work}
\label{sec:related-work}
\vspace{-1mm}

% In this section, we discuss relevant literature in developing and evaluating visualization recommendation algorithms.

%  \jane{Necessary to include ``Interface'' in the figure? Also, I wonder if we might try adding this figure to the paragraph text itself to reduce the amount of space it takes up.}

% \leilani{In general, the relationship between frameworks, algorithms, and systems is still not 100\% clear. I worry that if this is not clear, then we may get very low review scores simply because the reviewers do not understand how a system is different from an algorithm, and how an algorithm is different from a framework... A figure may seem wasteful, but it's not. I still strongly believe that a figure is the right way to go here to explain the relationships.}

\begin{figure}
\centering
 \includegraphics[width=0.6\columnwidth]{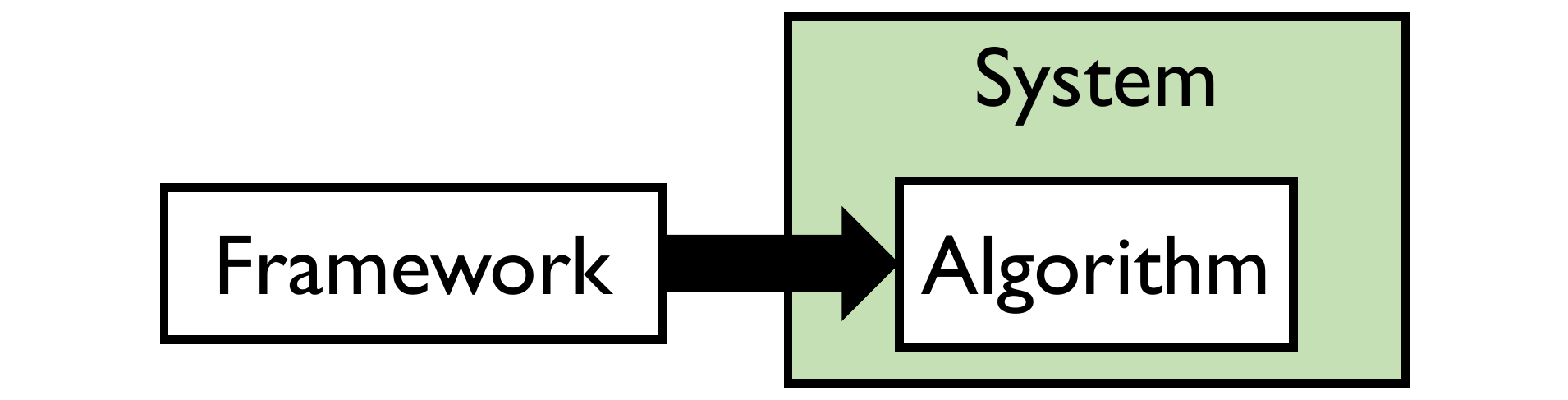}
 \vspace{-3mm}
 \caption{Visualization recommendation \emph{algorithms} are often specified using \emph{frameworks}, and evaluated using implemented \emph{systems}.}
 \label{fig:alg-sys-fwk}
 \vspace{-6mm}
\end{figure}

As illustrated in \autoref{fig:alg-sys-fwk}, users can only interact with visualization recommendation \emph{algorithms} when provided with an interface, and often a full \emph{system} through which to interact.
Furthermore, several algorithms are specified using existing visualization recommendation \emph{frameworks}, which often take the form of specialized languages.
%On the other hand, existing recommendation frameworks are proposed to model the design of recommendation algorithms,  or even further to guide the construction of new ones.
In this section, we discuss the relevant literature in specifying visualization recommendation algorithms, and evaluating both algorithms and systems. 

\vspace{-1mm}
\subsection{Visualization Recommendation Algorithms}
\vspace{-1mm}

Existing recommendation algorithms can be separated into two main categories based on how the ranking engine (oracle) is implemented: rule-based or machine learning-based.
Rule-based algorithms enumerate and then rank visualizations using heuristics based on theory or experimental findings in visual perception~\cite{Wongsuphasawat2015voyager, Wongsuphasawat2017voyager2,Mackinlay2007showme, Key2012vizdeck, Demiralp2017foresight, Seo2005rankbyfeature, Ehsan2016muve}. 
For example, theory work from Bertin~\cite{Bertin1983semiology} and Mackinlay~\cite{Mackinlay1986automating} has been incorporated within rule-based algorithms behind a wide range of recommendation systems, including Voyager~\cite{Wongsuphasawat2015voyager, Wongsuphasawat2017voyager2} and Show Me~\cite{Mackinlay2007showme}.
Other recommendation systems, such as VizDeck~\cite{Key2012vizdeck} and Foresight~\cite{Demiralp2017foresight} rank visualizations using manually-selected statistical rules.

Instead of ranking visualizations with manually-derived rules, other algorithms train machine learning models to generate recommendations~\cite{Hu2019vizml, Luo2018deepeye, Dibia2019data2vis,Moritz2018formalizing, Cao2020user, Wang2020datashot}.
For example, Hu et al.~\cite{Hu2019vizml} trained a deep learning model to learn the most common visualization designs from a large corpus of data sets and their associated Plotly visualizations.
%, and recommended visualization designs for new datasets using the trained model.
One of the Draco applications developed by Moritz et al.~\cite{Moritz2018formalizing}, Draco-Learn, was implemented by training models to learn effectiveness criteria from previous experimental findings.
% Neither Draco nor GraphScape is using machine-learning models.
% Dziban~\cite{Lin2020dziban}, which was built on top of Draco~\cite{Moritz2018formalizing} and Graphscape~\cite{Kim2017graphscape}, aims to output the most effective visualization with the shortest perceptual distance from a given visualization.
Luo et al.~\cite{Luo2018deepeye} strive to balance the best of both strategies by
combining deep learning with hand-written rules to generate recommendations.

Our framework provides a means of comparing these different ranking strategies, or oracles, in a systematic and repeatable manner.

\vspace{-1mm}
\subsection{Visualization Recommendation Frameworks}
\vspace{-1mm}

% There exists a gap between visualization design knowledge and automated visualization design applications.
% Thus, a couple of visualization design guidelines are proposed, which can be applied to visual design ranking during the recommendation process.

Several frameworks have been proposed to make it easier to create new visualization recommendation algorithms.
CompassQL~\cite{Wongsuphasawat2016towards} is a query language created by Wongsuphasawat~et~al., which can produce different types of recommendation algorithms by varying phases of the recommendation process, such as enumerating, choosing, and ranking.
For example, the visualization recommendation algorithm in Voyager~\cite{Wongsuphasawat2015voyager, Wongsuphasawat2017voyager2} is implemented with CompassQL.
ZQL~\cite{Siddiqui2017fast} is another query language that serves a similar purpose for visualization recommendation in the Zenvisage system.
% However, the contribution of both CompassQL and ZQL focuses on the ranking step, thus they are more like frameworks for oracles than frameworks for recommendation algorithms.
Draco~\cite{Moritz2018formalizing} is an alternative framework for specifying visualization recommendation algorithms based on answer set programming. Using Draco, one can specify new algorithms using a combination of encoding constraints and weights for these constraints. In this way, Draco enables the creation of new recommendation algorithms, without the creator having to worry about how to enumerate the underlying visualization design space. 
% /phoebe:  you use "alternative" twice in the above paragraph.  Might want to swap one of them out for a different word (unless alternative has additional significance)

To evaluate recommendation algorithms, we need to know not only the constraints imposed by the algorithms (the focus of current frameworks), but also the strategies employed to apply these constraints. Furthermore, we need to know what the differences are between strategies in order to reason about how they impact the performance of the system running \revised{the} algorithm and the decisions of users who view the recommendations. \revised{Current frameworks omit these details, making it difficult to use them to evaluate and compare different algorithms.}
%However, the aforesaid frameworks are designed for creating new recommendation algorithms, making it difficult to use them for the evaluation and comparison of different algorithms.
%For example, algorithms define a set of enumeration constraints or specifiers during each enumeration process, and visualizations are selected to the ranking step as long as they satisfy these constraints.
%Thus the relationship between candidate visualizations and the whole visualization design space is not clear, not to mention comparing the enumeration process among various algorithms.
In contrast, our framework gives a clear definition of the visualization design space and considers an algorithm's traversal method through enumeration, which makes the enumeration process comparable.

\vspace{-1mm}
\subsection{Evaluating Visualization Recommendation Algorithms}
\vspace{-1mm}
% \zehua{Just wrote this new subsection, please review.}

Evaluation is crucial since it \revised{provides evidence of whether a proposed algorithm} actually helps users explore their data more effectively.
However, not all algorithms are evaluated in terms of how they improve user exploration performance.
%However, evaluation is not always presented along with the proposal of recommendation algorithms or systems;
For example, Foresight~\cite{Demiralp2017foresight} only provides some usage scenarios to demonstrate its efficacy.
On the other hand, although some existing systems do empirically evaluate user performance, the proposed algorithms are evaluated in isolation.
% there are still no clues how the proposed algorithms benefit the actual analysis scenario compared to other algorithms.
% \leilani{The previous sentence is confusing. Is it just saying that even when user performance is measured, most algorithms are evaluated in isolation?}
% Even an evaluation was conducted, no comparison was presented or the comparison was not with other recommendation algorithms.
For instance, Voyager~\cite{Wongsuphasawat2015voyager, Wongsuphasawat2017voyager2} and SeeDB~\cite{Vartak2015seedb} were compared to a baseline with no recommendations provided.
VizDeck~\cite{Key2012vizdeck} claimed that \revised{VizDeck users completed} tasks with higher accuracy and less time compared to IBM ManyEyes~\cite{Viegas2007ManyEyes}, Google Fusion Tables~\cite{Gonzalez2010google}, and Tableau~\cite{Wesley2011analytic}.  However, none of the compared systems provide recommendations.
% neither the detailed evaluation process nor the analysis of benchmark results was present. 
% \phoebe is the baseline the same as "no recommendations provided", or are baseline and "no recommendations provided" separates things
% \leilani{ManyEyes, Google Fusion Tables, and Tableau do not provide recommendations... I'm pretty sure that's all we need to say here...}

% \leilani{visual perception *is* a form of human performance. This language is confusing. What we are really saying is that just because a user performs well on a low-level visual perception task does not mean that the user will perform equally well on a higher level task, such as prediction or data exploration. Most existing benchmarks for visualization recommendation optimize for low level perceptual metrics, which cannot be used to estimate user performance in real-world analysis tasks, which tend to be more complex. See my new argument at the end of \autoref{sec:examples}, the very last paragraph about evaluating the actual recommendations. It should help with this paragraph.}
%The evaluation and comparison of other recommendation algorithms, on the other hand, did not focus on the 
Even when multiple algorithms are compared in the literature, \revised{user performance} is still not the focus.
Dziban~\cite{Lin2020dziban} was evaluated by calculating the ranking of its recommended visualizations in both Draco~\cite{Moritz2018formalizing} and GraphScape~\cite{Kim2017graphscape} algorithms to check whether it provides a favorable tradeoff in terms of effectiveness and similarity. 
On the other hand, DeepEye~\cite{Luo2018deepeye} was tested by ground-truth data, which was derived by having students to label whether a visualization is good or bad.
Similarly, VizML~\cite{Hu2019vizml} was compared with CompassQL~\cite{Wongsuphasawat2016towards}, DeepEye~\cite{Luo2018deepeye}, ShowMe~\cite{Mackinlay2007showme} and Data2Vis~\cite{Dibia2019data2vis} using an effectiveness score which was calculated based on human-labeled data.
However, human-perceptually ``good'' visualizations do not necessarily help the actual analysis process.
Since tasks are not taken into account in the labelling process, there exists no evidence of whether these benchmark results can carry over into the actual human performance with higher level analysis tasks.
% Since the labeling process does not take tasks into considerations, there exists no evidence of whether these benchmark results can carry over into the actual human performance in practical analysis scenarios.

% Although Dziban~\cite{Lin2020dziban} presented a comparison with Draco~\cite{Moritz2018formalizing} and GraphScape~\cite{Kim2017graphscape}, on the other hand, VizML~\cite{Hu2019vizml} compares itself with CompassQL~\cite{Wongsuphasawat2016towards}, DeepEye~\cite{Luo2018deepeye}, ShowMe~\cite{Mackinlay2007showme} and Data2Vis~\cite{Dibia2019data2vis}, none of these comparisons focus on the user performance in the actual analysis scenario.
% Thus we have no clue of how these benchmark results could carry on to predict how users perform with different recommendation algorithms given various analysis tasks. 

\revised{In this paper, we show how our framework can be used to theoretically compare visualization recommendation algorithms, and empirically evaluate user performance for different visual analytics tasks.}

\vspace{-1mm}
\section{Evaluation Framework}
\label{sec:framework}
\vspace{-1mm}

% \leilani{We should have a table reflecting how big the design space is, and how big of a space each algorithm can actually cover. Gives a sense of total scope.}

%In this section, we describe the general recommendation process that the majority of visualization recommendation algorithms utilize.
In this section, we describe our framework, which is based on the general recommendation process followed by the majority of visualization recommendation algorithms: enumerate, search, and rank.
%on modeling visualization recommendation algorithms which follow the similar recommendation process. 
%We also provide a couple of examples of current existing recommendation systems that our framework can represent. 
To demonstrate how our framework can be applied, we compare five existing representative visualization recommendation systems: Voyager~\cite{Wongsuphasawat2015voyager, Wongsuphasawat2017voyager2}, DeepEye~\cite{Luo2018deepeye}, Foresight~\cite{Demiralp2017foresight}, Show Me~\cite{Mackinlay2007showme} and Dziban~\cite{Lin2020dziban}.
% Finally, we demonstrate that how our framework could be used to guide the construction of new recommendation algorithms.

\begin{figure}
\centering
 \includegraphics[width=1.0\columnwidth]{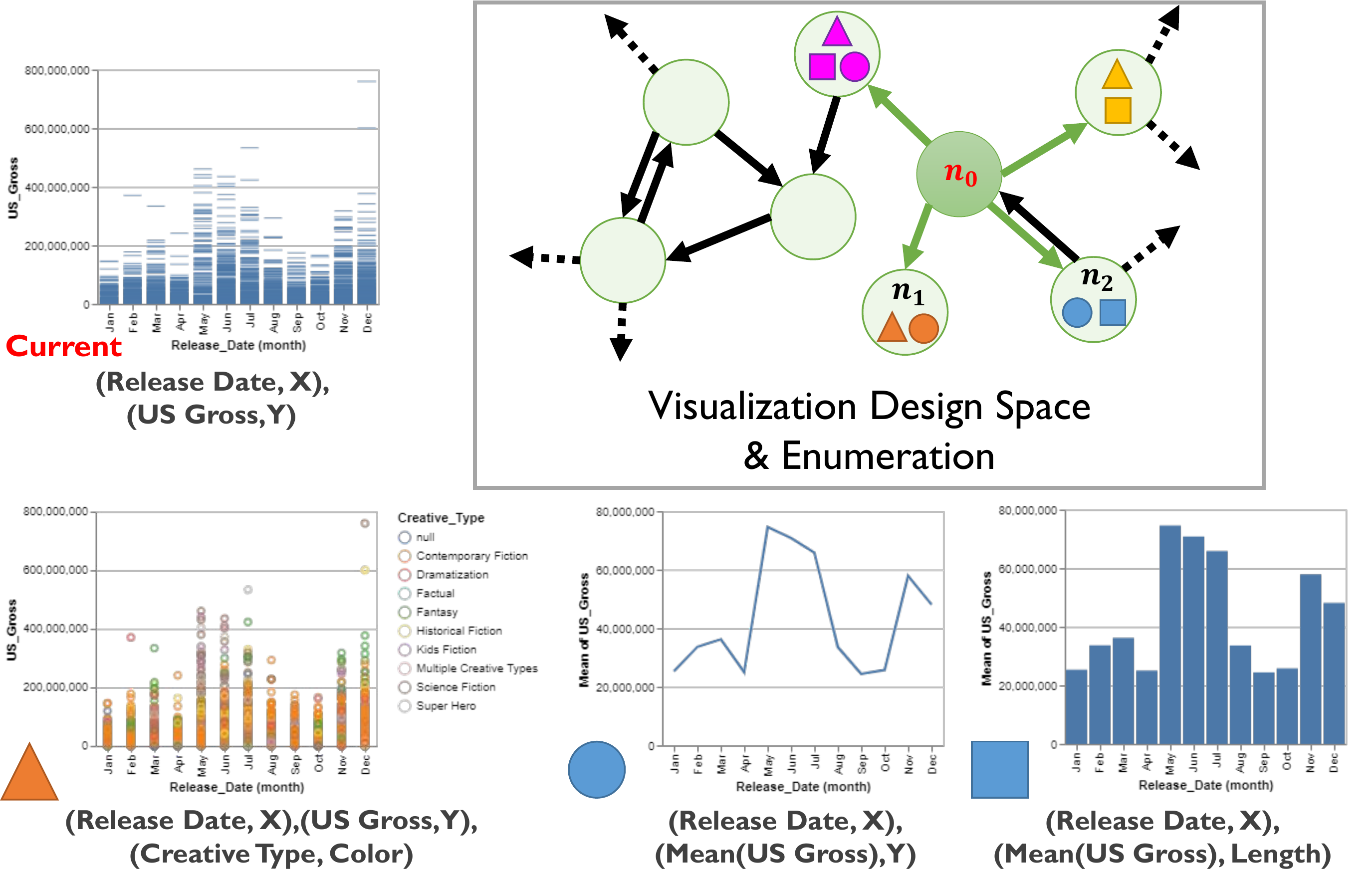}
 \vspace{-5mm}
 \caption{Illustration of the \textbf{visualization design space} and the \textbf{enumeration} step of a hybrid recommendation algorithm, using movies data as an example. The user's current visualization is at node $n_0$.}
 \label{fig:framework}
 \vspace{-6mm}
\end{figure}

% \leilani{Add to this figure, or make a subfigure, expanding node $n_2$ so we can see the encodings subgraph a bit}.

% \jane{What are the dark gray nodes in this figure that are not connected to anything? Why is the encoding for each attribute/transformation not considered part of the node (I have not gotten to the next section yet)? Does it make sense to discuss how this definition of the search space is different from other approaches (like GraphScape)}

% \leilani{Make two of the example visualizations be from the same node. Put the ``current'' visualization at the top, instead of that purple triangle one... The design space should probably also look like it's bigger (because it is, it's huge and not just 12 nodes...), I might include more edges that look like they go ``outside'' the example shown.}

\vspace{-1mm}
\subsection{Defining the Core Components of the Framework}
\label{sec:visualization-design-space}
\vspace{-1mm}
%Although various recommendation algorithms have been proposed and applied to visualization recommendation systems, the existing automated systems mostly follow a common recommendation process, which is first enumerating, then searching, finally ranking (see \autoref{fig:framework}).
Visualization recommendation algorithms are a form of search algorithm, which generally follow two basic steps: traverse candidates within the larger search space, and evaluate these candidates against specific search criteria. In the case of visualization recommendation, the traversal step involves \textbf{enumerating} the visualization design space, and the evaluation step requires \textbf{ranking} the candidate visualizations for the subsequent recommendation. However, before an algorithm can enumerate candidate visualizations, the \textbf{visualization design space} must first be clearly defined. In this section, we define the visualization design space, as well as the enumeration and ranking steps. 
% We illustrate each component in \autoref{fig:framework}.

\vspace{-1mm}
\subsubsection{Defining the Visualization Design Space}
\vspace{-1mm}
% \zehua{Rewrote based on Eunyee's and Jane's comments.}
% A distinguishing feature of our framework is how we take into account how different algorithms traverse the visualization design space. 

% \zehua{maybe we should specify how the visualization design space looks like in terms of three types of recommendations: data query, encoding, and hybrid (based on Eunyee's comments), then we can fix the showme problem, since showme is an encoding recommendation algorithm.}

To facilitate comparison, we must first establish a consistent definition of the visualization design space that can be applied to a wide range of algorithms.
% \revised{The design space is commonly modeled as a graph~\cite{Vartak2015seedb,Wongsuphasawat2016towards,Moritz2018formalizing,siddiqui2017fast}, however previous definitions only cover a fraction of the full design space.}
\revised{Prior work uses graph theory to model visualization spaces~\cite{Ma1999image, Jankun-Kelly2007model, Heer2008graphical}, however previous definitions cover only a fraction of the full design space~\cite{Vartak2015seedb, Wongsuphasawat2016towards, Moritz2018formalizing, Siddiqui2017fast}. We contribute a generalization of these existing visualization spaces using graph theory.
In our framework, we consider the \emph{full} design space of all possible visualizations,} which is defined as the combination of data attributes, encoding channels, and data transformations that can be applied to a given dataset. By leveraging this full design space, individual algorithms can be compared in terms of the particular subspaces they traverse.

\vspace{-1mm}
% \paragraph{Tracking Attributes and Data Transformations Within the Design Space Graph.}
\paragraph{Tracking Visualization Designs Within the Design Space Graph.}
% Suppose we are generating recommendations for a movies dataset $D$, containing attributes $a$ such as \textsl{movie title}, \textsl{creative type}, \textsl{gross}, \textsl{release date}, etc.
% \jane{Consider giving data attributes a unique style (e.g., \texttt{release date})}
Suppose we are generating recommendations for a movies dataset $D$, containing $n$ attributes $A = \{a_1, ..., a_n\}$, such as \textsl{movie title}, \textsl{creative type}, \textsl{gross}, \textsl{release date}, etc.
There are $m$ possible transformations $T = \{t_1, ..., t_m\}$; each transformation has a set of parameters to determine how it can be applied to the data. For example, one possible data transformation is calculating the average of movie gross: \texttt{AVG}($a_{gross}$), which is parameterized by only one attribute.
On the other hand, there also exist $k$ possible encoding channels $C = \{c_1, ..., c_k\}$ which are used to visualize the combination of attribute and data transformation, such as, the 13 encoding channels proposed by Mackinlay~\cite{Mackinlay1986automating}.
We represent the visualization design space for this dataset as a network graph $G = (N, E)$.
Each node of the graph $n \in N$ contains visualizations which are defined by a set of data attributes ($a_i$, $a_j$, etc.), data transformations ($t_i$, $t_j$, etc.), and encoding channels ($c_i$, $c_j$, etc.):

\vspace{-2mm}
\begin{equation}
    n = \{ v \ |\ v = [c_i(a_i, t_i), c_j(a_j, t_j), ... ] \} 
    % n = \{a_i, a_j, ..., t_i, t_j, ...\}
    %n = (A, T), A = [a_i, a_j, ...], T = [t_i, t_j, ...], t_i \in \widetilde{T}, t_j \in \widetilde{T}
\vspace{-2mm}
\end{equation}
%Here, $a_i$, $a_j$, etc. are attributes from $D$, while $t_i$, $t_j$, etc. are data transformations, or calculations applied to specific attributes, $a_i$, $a_j$, etc. 

% \noindent\jane{Is $t_i$ a transformation applied to $a_i$?  Consider a different numbering scheme that more clearly/formally describes the space, for example, consider updating the text following a form like: ``Suppose we are generating recommendations for a movies dataset D, containing $n$ attributes ($a_1, ..., a_n$)...'' Is there a way to more clearly/formally define transformations (e.g., how many are possible, how often can they be applied, what is the upper bound on this space)? Maybe something like: ``There are $m$ possible transformations $t_1, ..., t_m$; each transformation has a set of parameters to determine how it can be applied to the data. For example, one possible data transformation is calculating the average of movie gross: \texttt{AVG}($a_{gross}$), which is parameterized by only one attribute.'' Not sure this is quite there, but if we're using a formal specification like this, I would like some more robust discussion of how the space is defined.\par}

Here, $a_i$, etc. are attributes from $D$, and $t_i$, etc. are data transformations operated on attributes $a_i$, etc., while $c_i$, etc. are encoding channels used to visualize the combination of $(a_i, t_i)$, etc.
$v$ is a visualization defined by the attribute set $\{a_i, a_j\}$ and its corresponding data transformations $\{t_i, t_j\}$ and encoding channels $\{c_i, c_j\}$.
Edges between each pair of nodes represent operations that transform one node to another, such as adding one attribute, or changing the data transformation or encoding channel of an attribute. 

%For each node, if the attribute set, the data transformation added to each attribute, as well as the encoding channel visualizing each combination of attribute and data transformation are fixed, there exists only one visualization design in the node.
\vspace{-1mm}
\subsubsection{Defining Sub-spaces for Different Types of Algorithms}
\label{sec:subspaces}
\vspace{-1mm}

To more efficiently navigate the visualization design space,
recommendation algorithms can merge multiple visualizations into one node, or even ignore nodes, reducing the total edges that needed to be traversed.
We discuss how different algorithms manipulate the visualization design space, based on the three types of recommendation algorithms proposed by Wongsuphasawat et al.~\cite{Wongsuphasawat2016towards}:
%by adapting the category of recommendation algorithms proposed by Wongsuphasawat et al.~\cite{Wongsuphasawat2016towards}. 
%There are three types of recommendation algorithms based on whether the
algorithms suggesting what \textit{attributes and/or transformations} to visualize (data query recommendations), what \textit{encoding channels} to apply to selected data (visual encoding recommendations), or both (hybrid recommendations).

\vspace{-1mm} 
\paragraph{Visual Encoding Recommendations.}
These algorithms focus on enumerating and ranking variations in encoding choices (e.g., Show Me~\cite{Mackinlay2007showme}), requiring access to attribute, transformation, and encoding information.  However, to reduce the cost of enumerating the visualization design space, these algorithms often require the user to select what attributes and transformations to visualize in advance. In this way, all nodes that include \revised{non-user-selected} attributes can be ignored.
We can represent this \revised{user selection-based} subspace in the following way:
%Grouping visualizations with the same attribute set makes it easier to access a large group of different encoding designs at one time.
%The following equation shows the example of a node consists of visualizations with the same attribute set $(\bar{a_i}, \bar{a_j})$ but visualizing by different encoding channels:

%\vspace{-2.5mm}
%\begin{equation} \label{eq:node-visual-encoding}
%    n = \{ [c_1(\bar{a_i}), c_2(\bar{a_j})],..., [c_{k-1}(\bar{a_i}), c_k(\bar{a_j})] \}
%\vspace{-2.5mm}
%\end{equation}
\vspace{-2mm}
\begin{equation}
\label{eq:vis-encoding}
    n = \{ v  = [c_i(a_i, t_i), ... ]\ |\ \mathtt{selected}(a_i,t_i) = 1,\ \forall (a_i,t_i) \in v \} 
    % n = \{a_i, a_j, ..., t_i, t_j, ...\}
    %n = (A, T), A = [a_i, a_j, ...], T = [t_i, t_j, ...], t_i \in \widetilde{T}, t_j \in \widetilde{T}
\vspace{-2mm}
\end{equation}

\vspace{-1mm}
\paragraph{Data Query Recommendations.}
These algorithms tend to focus on recommending attributes and/or transformations, and ignore encoding channels (e.g., Foresight~\cite{Demiralp2017foresight} and SeeDB~\cite{Vartak2015seedb}).
% thus data transformations and encoding channels are little concerned.
%To access various attribute sets at one time,
%with the same combination of encoding channels visualizing different attribute sets are grouped in one node.
To ignore design variation, we can merge all visualizations that vary only by encodings into one node (i.e., remove all encoding channel specifiers $\{c_i, c_j, ...\}$):
%The example of a node containing visualizations
%using two position encoding channels (\texttt{X}, \texttt{Y}) to visualize all two-attribute pairs is as shown:

%\vspace{-2mm}
%\begin{equation} \label{eq:node-data-query}
%    n = \{ [\bar{c_\texttt{X}}(a_1), \bar{c_\texttt{Y}}(a_2)],..., [\bar{c_\texttt{X}}(a_{n-1}), \bar{c_\texttt{Y}}(a_n)] \}
%\vspace{-2mm}
%\end{equation}

\vspace{-2mm}
\begin{equation}
    n = \{ v \ |\ v = [(a_i, t_i), (a_j, t_j), ... ] \} 
    % n = \{a_i, a_j, ..., t_i, t_j, ...\}
    %n = (A, T), A = [a_i, a_j, ...], T = [t_i, t_j, ...], t_i \in \widetilde{T}, t_j \in \widetilde{T}
\vspace{-2mm}
\end{equation}

\vspace{-1mm}
\paragraph{Hybrid Recommendations.}
These algorithms consider variations in attributes, transformations, and encoding channels (e.g., Voyager~\cite{Wongsuphasawat2015voyager,Wongsuphasawat2017voyager2}). As a result, the full expressiveness of the visualization design space graph is required. However, enumerating all possible combinations of attributes, transformations, and encoding channels can be prohibitively expensive. In the next section, we discuss how algorithms efficiently enumerate candidate visualizations within this space of possible visualization designs.

\vspace{-1mm}
\subsubsection{Defining the Enumeration Step}
\label{sec:enumeration}
\vspace{-1mm}
% \zehua{Modified based on Jane's comments.}

\noindent Given a formal definition of the visualization search space $G$, search algorithms must then traverse this space to identify qualified candidate results.
The result of the enumeration step is a list of candidates that match the input requirements, which are then passed to the ranking step.
However, the full visualization design space is an exponential function of attributes $\{a_1, ..., a_n\}$, transformations $\{t_1, ..., t_m\}$, and encoding channels $\{c_1, ..., c_k\}$, making it prohibitively large to search in its entirety. As a result, visualization recommendation algorithms must address a trade-off between recommendation breadth and execution cost, where higher quality results can be achieved by enumerating and ranking more of the visualization design space, but performing this additional work increases the algorithm's execution time.
%Enumeration is the first step of the recommendation process.
%In this step, visualization recommendation systems would keep all possible output candidates in the space.
%It is the role of the enumeration step to efficiently traverse the visualization design space, where the goal is
%%During traversal, the goal of the visualization recommendation algorithm is
%to identify potential candidates with minimal effort while preserving recommendation quality.

\vspace{-1mm}
\paragraph{Input Nodes to the Enumeration Step.}
%Arbitrarily searching $G$ can be expensive.
In response to this tradeoff, recommendation algorithms generally enumerate visualizations based on one or more reference nodes, often the nodes that contain the user's current selected attributes or visualization, or auto-generated reference nodes derived from simple heuristics. 
For example, Voyager~\cite{Wongsuphasawat2015voyager,Wongsuphasawat2017voyager2} uses the node that contains the user's current visualization as a reference (denoted as $n_0$), otherwise Voyager generates univariate visualizations by default. 
% For example, Voyager~\cite{Wongsuphasawat2015voyager,Wongsuphasawat2017voyager2} uses the user's current visualization $n_0$ as a reference, otherwise Voyager generates univariate visualizations by default. 
% In another example, SeeDB~\cite{Vartak2015seedb} generates two-attribute bar charts for reference, which are later compared to bar charts containing an additional group-by transformation.

\vspace{-1mm}
%\paragraph{Applying Constraints to Bound the Total Nodes to be Enumerated.}
\paragraph{Applying Constraints to Bound the Number of Candidate Nodes.}
\revised{Algorithm designers tend to keep the space of traversed visualizations quite small} by imposing strict manual constraints on what parts of the space can be traversed.
The most common constraints limit either the maximum path length that can be explored from some reference node $n_0$, or the maximum number of inputs \revised{contained} within a candidate node. Using our design space notation from \ref{sec:visualization-design-space}, we can represent all nodes with a maximum path length of 2 from $n_0$ as:
\vspace{-2mm}
\begin{equation}\label{eq:path-constraint}
    \{ n\ |\ \texttt{dist}(n, n_0)\ \leq 2,\ n \in N\},
    \vspace{-2mm}
\end{equation}
and nodes comprised of visualizations ($v$) with at most two inputs as:
\vspace{-2mm}
\begin{equation}\label{eq:input-constraint}
    \{ n \ \mid{}\ |v| \leq 2 ,\ v \in n ,\ n \in N\}.
    % \{ |A| \leq 2\ , |\widetilde{T}| \leq 2\  \mid{}\ n = (A, T) \in N \}.
    \vspace{-2mm}
\end{equation}
For example, Voyager~\cite{Wongsuphasawat2015voyager, Wongsuphasawat2017voyager2} only considers nodes that differ from the user's current visualization by at most one attribute or data transformation, i.e., by setting the path length threshold to one for \autoref{eq:path-constraint}. This example is illustrated in \autoref{fig:framework}.
% DeepEye~\cite{Luo2018deepeye} limits the number of inputs that can be represented within input nodes to two data attributes, 
% and at most two data transformations (aggregating and binning)
% i.e., by setting $|A| = 2$ in \autoref{eq:input-constraint}.
DeepEye~\cite{Luo2018deepeye} only outputs two-attribute visualizations, 
i.e., by setting $|v| = 2$ in \autoref{eq:input-constraint}.
DeepEye also limits data transformation choices to three types (aggregating, binning and sorting), and encoding choices to one of four basic visualization types (bar, pie, line, scatterplot), i.e., by setting $|T| = 3$ and $|C| = 4$.

\vspace{-1mm}
\paragraph{Navigating the Bounded Design Space to Enumerate Candidates.}
Once the constraints of the traversal are established, then algorithms must select a method for enumerating specific designs within this bounded space. Given one or more reference nodes, there are three basic approaches to the enumeration process:
\begin{itemize}[nosep]
    \item a \textbf{random traversal}, such as by listing random combinations of valid attributes, transformations, and/or encoding channels;
    \item a \textbf{tree-oriented traversal}, such as breadth-first or depth-first search along $G$, originating at $n_0$;
    \item a \textbf{cluster-oriented traversal}, where nodes are clustered by pre-defined criteria, and clusters closest to $n_0$ are prioritized.
\end{itemize}
%Enumeration may also occur in an online fashion, allowing the enumeration step to stop early if candidates of sufficient quality are identified by the ranking step. 

% \noindent\jane{I feel like the ``cluster-oriented traversal'' is another argument for \emph{not} pre-clustering the graph based on encodings; instead, that could be something done by a specific approach.\par}

We see that these traversal strategies lead to varying degrees of depth and breadth in the coverage of the design space. For example, random and cluster-oriented traversals can cover a broader range of $G$, but at the risk of having few nodes explored close to the user's current visualization $n_0$. In contrast, tree-oriented traversals will have dense coverage near $n_0$, but may have little or no coverage elsewhere in $G$.

Note also that this traversal process need not take place all at once. For example, in the case of algorithms that rely on machine learning models, enumeration may happen both in the training phase (random traversal of training inputs) as well as in the prediction phase (cluster-oriented traversal within the model structure).

\vspace{-1mm}
\subsubsection{Defining the Ranking Step}
\label{sec:ranking}
\vspace{-1mm}

% \jane{I feel like the fact that we have to say things like ``including encodings'' really emphasizes the limitation of the current design space definition, so it might be good to discuss alternatives.}

\noindent Given the candidates generated by the enumeration step, the purpose of the ranking step is to order these candidates in terms of how closely they match a set of pre-defined search criteria. In the case of visualization recommendation algorithms, the search criteria represent the quality and relevance of the candidate visualization.
%An example is illustrated in \autoref{fig:framework}, Ranking.
%Oftentimes after the search process, there are still a large number of visualization candidates remaining.
%The visualization recommendation system then needs a ranking engine to suggest the optimal candidate or the top candidates (\autoref{fig:framework}, Ranking).
We use 
% the term
``oracle'' to refer to the part of the algorithm that assesses candidate quality and relevance.

\vspace{-1mm}
\paragraph{Oracle Inputs \& Structure.}
Oracles often take as input the user's recent history of visualizations created and interactions performed, as well as statistics about the current dataset. Using these inputs, oracles typically compute one or more scoring features and rank enumerated candidates using a weighted function of these features, or a model. Feature weights for the model can be represented mathematically, such as by assigning numerical weights to calculated heuristics to produce a single score~\cite{Lin2020dziban,Wongsuphasawat2015voyager,Wongsuphasawat2017voyager2}, as well as procedurally, such as through ordered pruning rules to eliminate low-quality candidates~\cite{Luo2018deepeye}.
There are three types of models that oracles often use to rank candidates: behavioral models, statistical models, and machine learning models. 

\vspace{-1mm}
\paragraph{Behavioral models.} These models are generally represented as manual heuristics derived from user studies and/or field observations. For example,  APT~\cite{Mackinlay1986automating}, Draco-APT~\cite{Moritz2018formalizing}, Show Me~\cite{Mackinlay2007showme}, and Voyager~\cite{Wongsuphasawat2015voyager,Wongsuphasawat2017voyager2} are based in part on manually-derived best practices, particularly for enhancing visual perception. In another example, the BDVR algorithm~\cite{gotz2009behavior} compares the user's most recent interactions to the four most common interaction patterns observed with the HARVEST system. The BDVR algorithm then ranks visualizations based on whether they would be produced by the closest matching patterns.

\vspace{-1mm}
\paragraph{Statistical models.}
These models often use a pre-defined set of aggregate statistics to compare candidates~\cite{Demiralp2017foresight, Key2012vizdeck, Vartak2015seedb}. For example, Foresight~\cite{Demiralp2017foresight} analyzes the dataset to be visualized for statistical properties selected by the user, such as skew, outliers, and linear relationships, and scores candidate visualizations according to these features.

\vspace{-1mm}
\paragraph{Machine Learning models.}
These models take large corpora of existing user data as input to an offline training phase~\cite{Moritz2018formalizing}.
During the training phase, these models generally cluster similar visualization designs, and develop hierarchical data structures to efficiently index into these clusters.
Recent approaches use deep learning models to avoid the need for feature engineering prior to training~\cite{Luo2018deepeye,Hu2019vizml,Dibia2019data2vis}.
%Data query recommendation systems might calculate the statistics of the candidate variable sets~\cite{ Demiralp2017foresight, Key2012vizdeck, Vartak2015seedb}, while encoding recommendation systems might calculate the effectiveness of the candidate encodings~\cite{Mackinlay2007showme, Wongsuphasawat2015voyager, Wongsuphasawat2017voyager2}.
%~\cite{Mackinlay2007showme, Wongsuphasawat2015voyager, Wongsuphasawat2017voyager2}

\vspace{-1mm}
\paragraph{Hybrid models.}
Hybrid oracles are also possible, where multiple models may be used.
%For hybrid systems, a ranking engine which considers both data query and encoding parameters is needed.
Oracles may also need to prune redundant candidates if they are too similar in quality and relevance~\cite{Mackinlay2007showme, Wongsuphasawat2015voyager, Wongsuphasawat2017voyager2}.

\vspace{-1mm}
\subsection{Comparing Existing Algorithms Using the Framework}
\label{sec:examples}
\vspace{-1mm}

\begin{figure}
    \begin{subfigure}[t]{0.3\columnwidth}
        \vskip 0pt
        \centering
        \includegraphics[width=\textwidth]{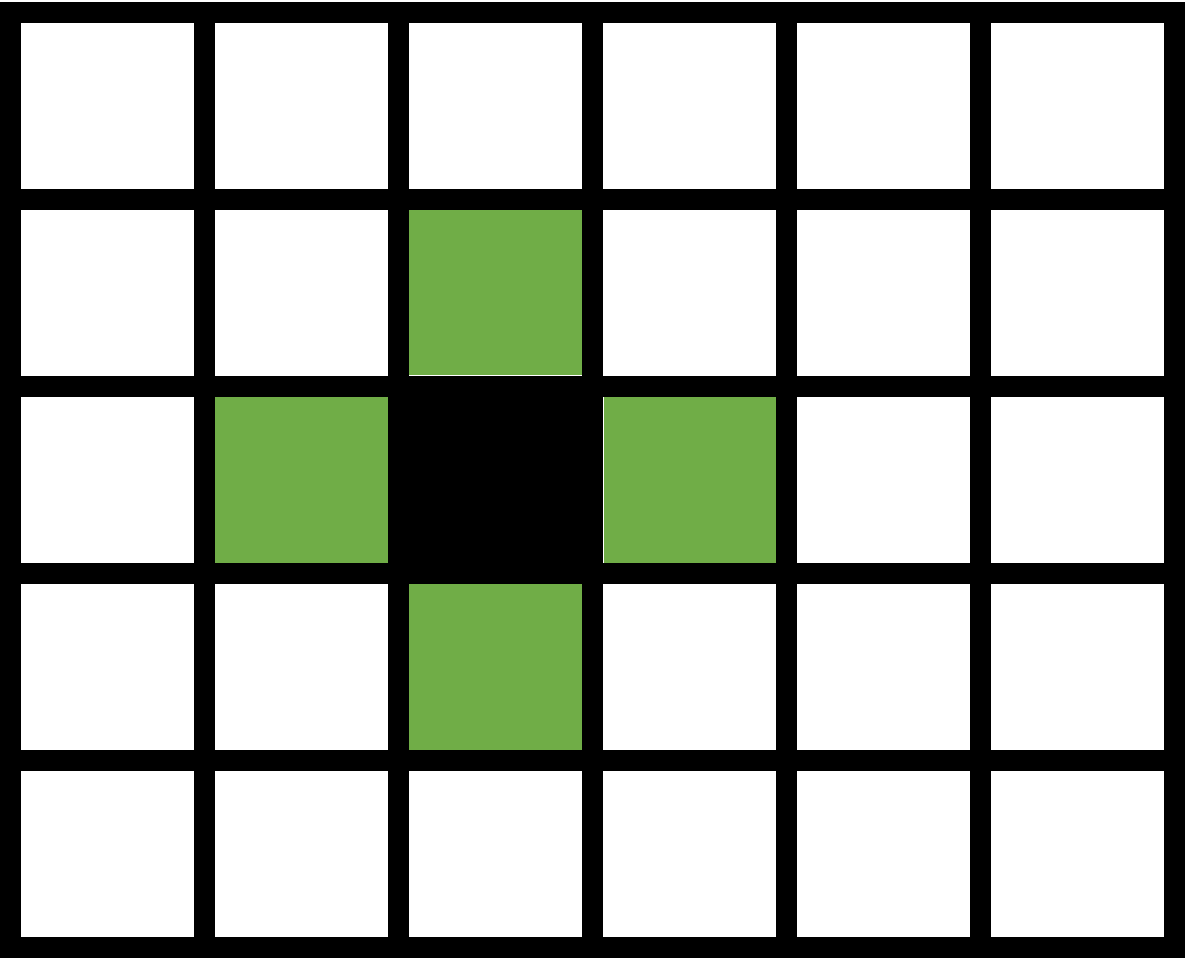}\vspace{-2mm}
        \caption{Voyager~\cite{Wongsuphasawat2015voyager, Wongsuphasawat2017voyager2} uses tree-based enumeration with max path length of 1.}
        \label{fig:ex-voyager2}
    \end{subfigure}\hfill%
    \begin{subfigure}[t]{0.3\columnwidth}
        \vskip 0pt
        \centering
        \includegraphics[width=\textwidth]{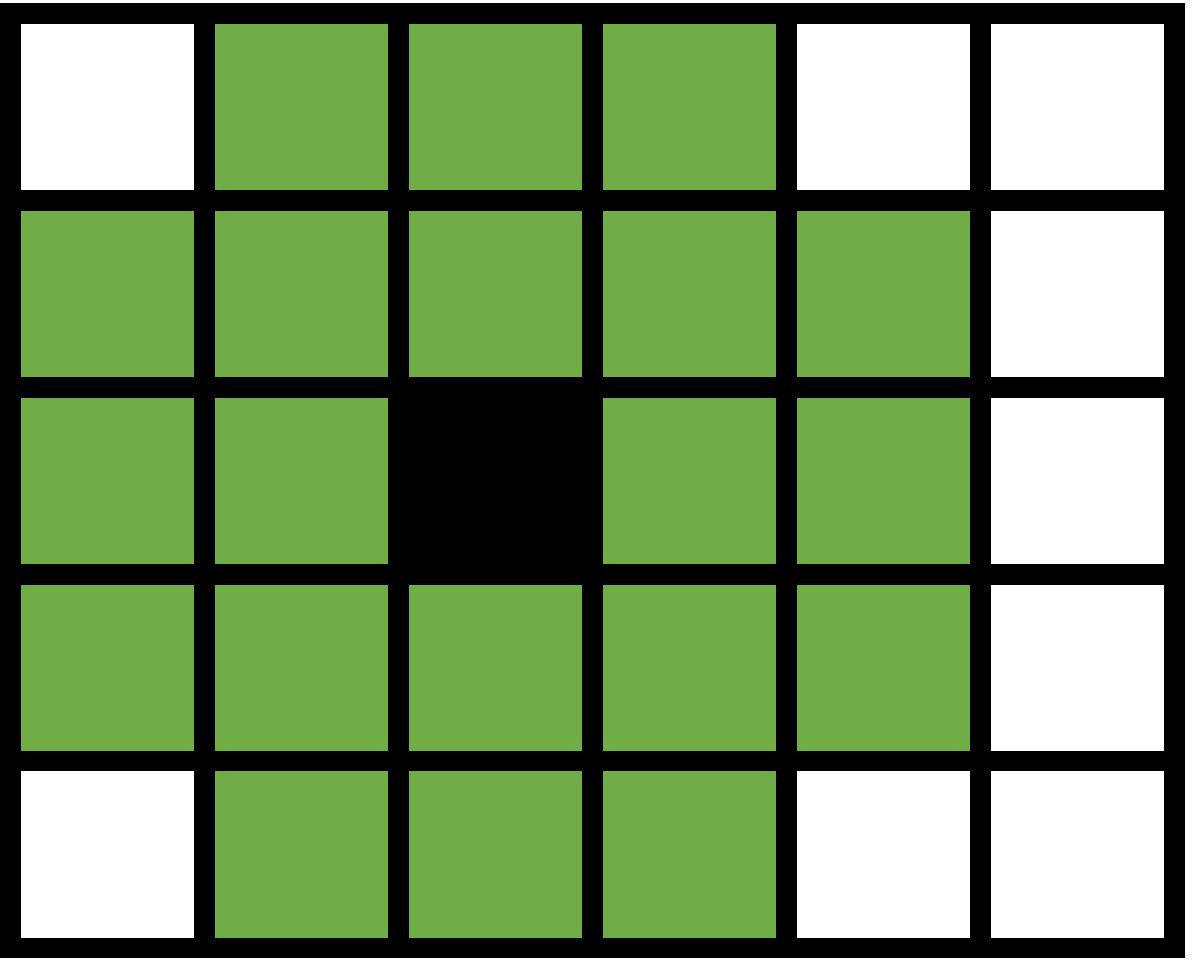}\vspace{-2mm}
        \caption{Foresight~\cite{Demiralp2017foresight} enumerates all data attributes with a max input of 2.}
        \label{fig:ex-foresight2}
    \end{subfigure}\hfill%
    \begin{subfigure}[t]{0.3\columnwidth}
        \vskip 0pt
        \centering
        \includegraphics[width=\textwidth]{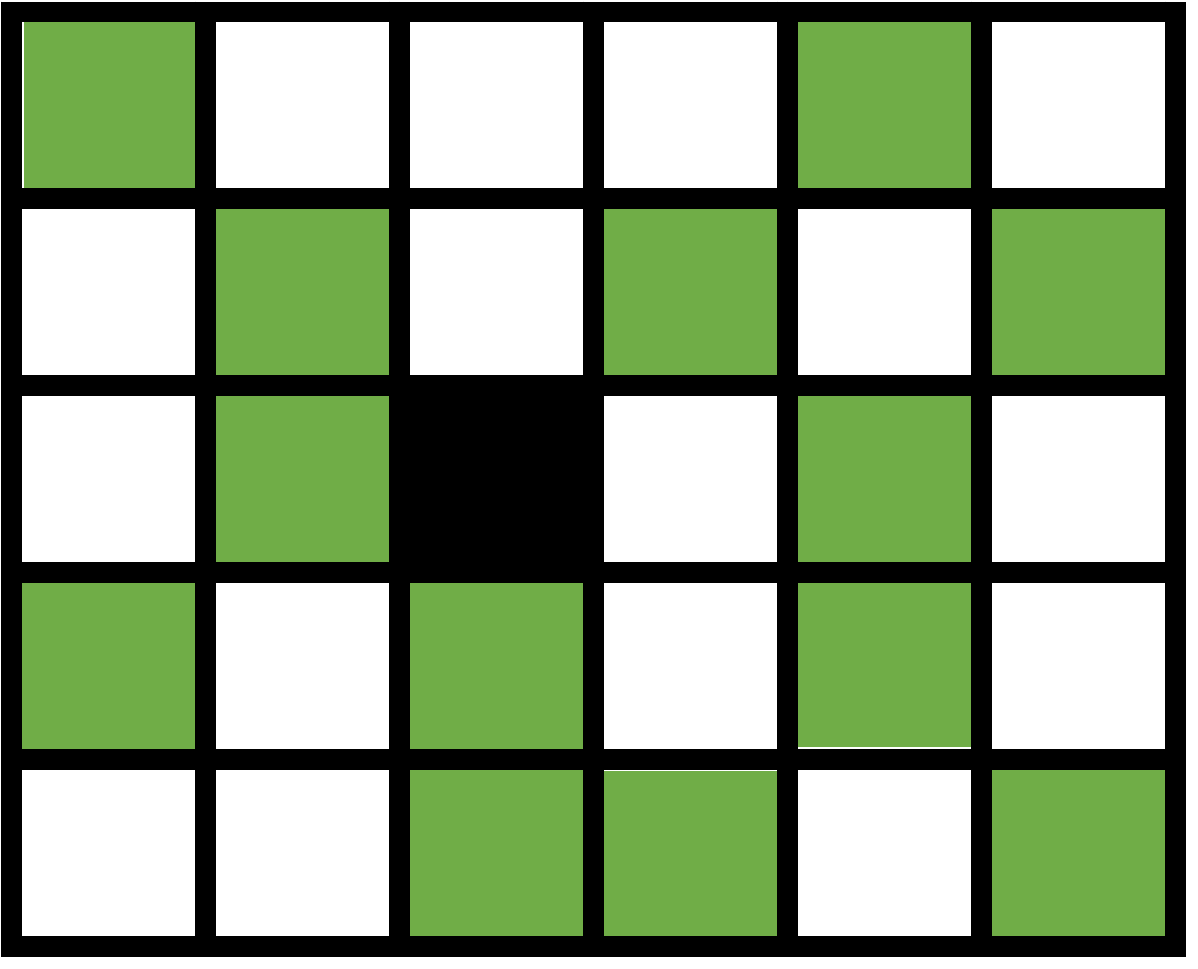}\vspace{-2mm}
        \caption{DeepEye~\cite{Luo2018deepeye} uses a clustered enumeration with an input of 2 attributes.}
        \label{fig:ex-deepeye2}
    \end{subfigure}\hfill%
    
    % \begin{subfigure}[b]{0.7\textwidth}
    % \centering
% \begin{tabular}{llll}
%     &   & \multicolumn{2}{l}{\textbf{Training Cost}} \\
%     &   & Low & High  \\ \cline{3-4} 
% \multirow{2}{*}{\textbf{\begin{tabular}[c]{@{}l@{}}Ranking\\ Cost\end{tabular}}} & \multicolumn{1}{l|}{Low}  & \multicolumn{1}{l|}{Voyager}                                                     & \multicolumn{1}{l|}{}        \\ \cline{3-4} 
%  & \multicolumn{1}{l|}{High} & \multicolumn{1}{l|}{\begin{tabular}[c]{@{}l@{}}Dziban,\\ Foresight\end{tabular}} & \multicolumn{1}{l|}{DeepEye} \\ \cline{3-4} 
% \end{tabular}
%     \caption{We bin the oracles for Voyager, Dziban, Foresight, and DeeyEye according to the effort required to train the corresponding models (training cost), and the costs incurred when ranking the candidates (ranking cost). \leilani{Cut d?}}
%         \label{fig:ranking-table}
    % \end{subfigure}
    \vspace{-2mm}
 \caption{A comparison of attribute enumeration methods for three existing recommendation algorithms. Each square is a node in the visualization design space, where the current node ($n_0$) is colored black.}
 \label{fig:examples2}
 \vspace{-3mm}
\end{figure}

Using the three main components of our framework, we can evaluate a wide range of visualization recommendation algorithms.
We demonstrate the flexibility of our framework by analyzing algorithms from five existing works: Voyager~\cite{Wongsuphasawat2015voyager, Wongsuphasawat2017voyager2}, DeepEye~\cite{Luo2018deepeye}, Foresight~\cite{Demiralp2017foresight}, Show Me~\cite{Mackinlay2007showme} and Dziban~\cite{Lin2020dziban}.
We compare the high-level intuition behind the enumeration strategies in \autoref{fig:examples2}, and the enumeration constraints in \autoref{tab:constraints}.
We selected these five algorithms because they cover all three types of recommendation algorithms proposed by Wongsuphasawat et al.~\cite{Wongsuphasawat2016towards}, and their results can be generalized to many other systems.
%Here, we use Voyagers~\cite{Wongsuphasawat2015voyager, Wongsuphasawat2017voyager2}, Foresight~\cite{Demiralp2017foresight}, and DeepEye~\cite{Luo2018deepeye} as the examples, since extending them to other automated systems is easy.
For instance,
%Voyager~\cite{Wongsuphasawat2015voyager, Wongsuphasawat2017voyager2} and DeepEye~\cite{Luo2018deepeye} are hybrid algorithms, where Voyager utilizes a rule-based ranking model, and 
Show Me~\cite{Mackinlay2007showme} uses a visual encoding recommendation algorithm.
Foresight uses a query recommendation algorithm that is similar to other query recommendation algorithms (e.g., VizDeck~\cite{Key2012vizdeck}).
%As data query recommendation algorithms, VizDeck~\cite{Key2012vizdeck} and Foresight~\cite{Demiralp2017foresight} are similar, differing mainly in the particular statistics they use to rank designs.
%Instead, they are using different training dataset or different machine learning models to train the ranking engine~\cite{Hu2019vizml, Moritz2018formalizing}.
%Note that no one algorithm is necessarily the best over all others, since they all make different trade-offs in enumeration and ranking performance.
DeepEye uses a hybrid algorithm that is machine learning based~\cite{Luo2018deepeye}.
Other machine learning approaches are similar to DeepEye, differing primarily by model type or input data used for training~\cite{Hu2019vizml, Moritz2018formalizing}.

\begin{table}
\caption{A comparison of enumeration constraints for existing visualization recommendation algorithms.}
\vspace{-3mm}
\centering
\begin{tabular}{C{5.5em}|C{5em}C{6.5em}C{5.5em}}
\toprule
\small \textbf{Algorithms} & \small \textbf{\# of Attributes} & \small \textbf{Transformations} & \small \textbf{Encodings} \\
\midrule
\small Voyager, Dziban, \newline \small Show Me & \small N/A & \small aggregation, binning, sorting & \small  position, length, area, shape, color\\
\rowcolor[HTML]{ededed} 
\small Foresight & \small $\leq 2$ & \small N/A & \small position, length\\
\small DeepEye & \small $=2$ & \small aggregation, binning, sorting & \small position, length, area\\
\bottomrule 
\end{tabular}
\label{tab:constraints}
\vspace{-5mm}
\end{table}

\vspace{-1mm}
\paragraph{Show Me~\cite{Mackinlay2007showme}.} 
As discussed in \autoref{sec:subspaces}, Show Me recommends visual encodings based on user-specified attributes and data transformations.
By assuming the attributes and transformations are fixed, Show Me can enumerate and rank all relevant nodes that vary only by visual encodings (see \autoref{eq:vis-encoding}).
%If Show Me groups all visualizations with the same attribute set and data transformations into one node (as illustrated in \autoref{eq:node-hybrid}), no traversal approach is needed.
%The algorithm could just visit the specific node ($n_0$) with the same user-specified attributes and data transformations, and then rank all visualizations to provide the ordered list of optimal visual encodings.
%It is a variant of our framework that the traversal method is not needed when all candidate visualizations are grouped into a single node.
% It is a limitation of our framework that ``traversing'' is not needed when all candidate visualizations are all grouped into a single node.
% It is a limitation of our framework that ``traversing'' encodings for a certain attribute/transformation set (i.e., a specific $n$) is not possible because all candidate visualizations are all grouped into a single node.

\vspace{-1mm}
\paragraph{Voyager~\cite{Wongsuphasawat2015voyager, Wongsuphasawat2017voyager2}.}
Voyager applies tree-oriented enumeration with an aggressively bounded search space in terms of attributes.
As mentioned in \ref{sec:enumeration}, Voyager uses the user's current visualization to generate relevant charts with a maximum path length constraint of one (see \autoref{fig:ex-voyager2}).
Constraining the attribute space allows Voyager to enumerate more encoding channels than other algorithms, as shown in \autoref{tab:constraints}.
%, such as \textsl{position}, \textsl{length}, \textsl{area}, \textsl{shape}, and \textsl{color}.
%The specified view in Voyager systems is generated based on the data fields or the chart selected by users.
%To help the data exploration, the Voyager systems generate recommended charts in the related views according to the specified chart.
%As illustrated in \autoref{fig:ex-voyager2}, the \textbf{visualization design space} for Voyager systems is a directed cyclic graph where each node represents a set of visualizations, and the directed edge connects two nodes represent the data or design transformation action from one to another.
%The action could be adding data variation to the current visualization, like adding an extra attribute or changing the data transformation (binning or averaging), while it could also be adding design variation like changing visual encodings.
%The related views are generated by applying one action to the current visualization.
%For example, as show in \autoref{fig:ex-voyager}, if the current specified chart is in the yellow node, then Voyager would traverse nodes in orange to search for candidates for the corresponding related views. 
%Thus, the \textbf{graph traversal method} that Voyager uses is breadth-first search.
%The Voyager \textbf{oracle} for ranking these neighbor visualizations is implemented using CompassQL~\cite{Wongsuphasawat2016towards}.
%Although CompassQL is a query language which could express various ranking rules,
The Voyager oracle applies Mackinlay's effectiveness rules~\cite{Mackinlay1986automating}.
%, implemented using CompassQL~\cite{Wongsuphasawat2016towards}.

% \noindent\jane{The voyager section does not discuss the possible encodings like the others.\par}

\vspace{-1mm}
\paragraph{Dziban~\cite{Lin2020dziban}.} 
Dziban is a visualization recommendation API that uses Draco~\cite{Moritz2018formalizing} as the implementation base. 
%Dziban supports both ``cold'' recommendations, or recommendations with no reference visualization, as well as ``anchored'' recommendations (i.e., with a specified $n_0$).
Dziban contains a hybrid visualization recommendation algorithm that builds on the GraphScape~\cite{Kim2017graphscape} and Voyager~\cite{Wongsuphasawat2015voyager,Wongsuphasawat2017voyager2} oracles. Given a user's prior query, Dziban can recommend new transformations and encoding channels, however, it does not recommend new attributes to visualize.
% (see \autoref{tab:constraints}).
%Dziban is a visual encoding recommendation algorithm as it recommends the optimal encoding channels to user-specified attributes.
Dziban prioritizes perceptually similar visualizations in its ranking step.
%It recommends the most effective encoding channels for the selected attributes when the reference is not given, on the other hand, when the specified visualization exists, it recommends the chart that provides the ``best'' tradeoff between effectiveness and similarity to the specified visualization.
%Similar to Show Me, Dziban only ``traverses'' inside the current node, however, Dziban also consider data transformation alternatives, thus the data transformation for the attribute set is not fixed in each node (i.e., grouping all possible visualization designs with the same attribute set in one node).
%Same as Voyager, Dziban also supports a large range of data transformations and encoding channels.
%Dziban allows for partial visualization specifications, which translates in our framework to a bounded search space with a maximum input size of four for \autoref{eq:input-constraint}. 
% Dziban also allows for variation in encoding channels.
% Through Draco, Dziban leverages a constraint solver to identify and rank candidate visualizations, where constraint solvers prune the search space of nodes that violate the specified constraints.

% \leilani{use latex paragraph commands, not textbf commands.}
\vspace{-1mm}
\paragraph{Foresight~\cite{Demiralp2017foresight}.}
The Foresight system ranks visualizations based on ``insight'' scores derived from user-selected statistical features or data attributes.
%Although Foresight is a data query recommendation system, it would still generate visualizations as the result of the recommendation.
Foresight enumerates all possible pairings of data attributes, as well as all individual attributes, but restricts the final visualizations to either a bar chart, a box plot, or a scatterplot.
Thus, Foresight performs a full attribute enumeration within a bounded search space.
%As show in \autoref{fig:ex-foresight}, the \textbf{visualization search space} for the Foresight system is a tree where the first level consists of visual designs with 1-attribute while the second level is composed of visualizations with 2-attribute.
%Different color of nodes represent different visualization types.
%Foresight recommends charts based on the user input.
%and ranks all enumerated combinations based on user-selected statistical features.
%, such as dispersion, skew, heavy tails, outliers, heterogeneous frequencies, etc.
%On the other hand, Foresight also allows user to fix a data attribute $\overline{x}$, then ranks statistical scores over pairs of $(\overline{x}, y)$.
%Since Foresight would traverse more than one levels to search for visualization candidates, thus we consider its \textbf{graph traversal method} as depth-first search.
% User can either pick a statistical feature or a variable to see which variable pairs have the highest \textit{insight} score.
% Since Foresight would traverse down to the second level to search for visualization candidates, thus we consider its \textbf{graph traversal method} as depth-first search.
%The \textbf{oracle} ranks the visualizations is the formulas for calculating the statistical \textit{insight} scores. 
% Users can select one or more statistical features as to calculate the ranking of data attribute tuples, including dispersion, skew, heavy tails, outliers, heterogeneous frequencies, etc.

\vspace{-1mm}
\paragraph{DeepEye~\cite{Luo2018deepeye}.}
Although DeepEye can be extended to support different numbers of attributes, the paper focuses on enumerating visualizations with two attributes and at most three data transformations (see \autoref{tab:constraints}). 
DeepEye supports four visualization types: bar, pie, line, and scatter.
%, as well as two data transformations, binning and aggregation.
%DeepEye can also decide if to order one of the attributes or neither of them.
Though the DeepEye authors describe their enumeration method in terms of trees, when compared using our evaluation framework, DeepEye actually performs cluster-oriented enumeration.
% DeepEye computes a score for every qualified node, thus the \textbf{graph traversal method} is depth-first search.
The oracle ranks visualization candidates using both hand-written rules from visualization experts, and a suite of binary classifiers trained using visualization preference data collected from user studies. Note that the hand-written rules are used as heuristics to prune the search space, interleaving the enumeration and ranking steps.

%  use a range of complex techniques to
%When compared using our evaluation framework, we find that Dziban is actually primarily a ranking engine, and not a full recommendation algorithm, since it lacks a means of enumerating the visualization design space. Similar to Voyager, Dziban also uses visual perception principles to rank candidate visualizations.

% \leilani{Add Show Me here. Briefly mention limitations of our framework.}

% \noindent\jane{As I've mentioned in previous comments, I feel like it is still somewhat hard to imagine or compare how these systems fit into the existing framework at a glance. I wish there was a concise mathematical representation and/or clear visualization of the results that we could include here to facilitate the comparison described in the next two sections. \par}

\vspace{-1mm}
\paragraph{Comparing Algorithms in Terms of Enumeration Trade-Offs.} We see wide variation in the depth and breadth of design space coverage in \autoref{fig:examples2}, and also in the enumeration constraints in \autoref{tab:constraints}.
For example, \revised{Voyager provides broad attribute and transformation coverage} near the user's current visualization, represented in black in \autoref{fig:ex-voyager2}, but Voyager leaves much of the visualization design space unexplored. 
However, Voyager enumerates more encoding channels compared to other algorithms, as shown in \autoref{tab:constraints}.
Dziban does not enumerate attributes, limiting its search space to transformations and encoding channel variations only; in return, Dziban can also enumerate a larger range of encoding channels.
Show Me takes this restriction one step further by only enumerating and ranking variations in encoding channels, enabling broad and deep coverage of the encoding space, but virtually no coverage of the attribute and transformation space.

In comparison, we see in \autoref{fig:ex-foresight2} that Foresight enumerates all attribute combinations within its bounded search space, providing both broader and deeper coverage of attributes.
However as a trade-off, we see in \autoref{tab:constraints} that Foresight severely limits the space of encoding channels that may be enumerated.
%Although at the expense of visiting many more nodes, each node does not contain as many visualization designs (no transformations, less encodings) as other algorithms do.
We see that DeepEye makes a similar tradeoff to Foresight.
%between the enumeration constraint and the coverage of the search space.
In \autoref{fig:ex-deepeye2}, we see that DeepEye's cluster-oriented enumeration approach provides greater enumeration depth than both Voyager and Foresight, but it also lacks thorough coverage of attributes (and transformations) across the bounded search space.
%On the other hand, DeepEye provides a fair amount of transformation and encoding choices.
However, the cost of this increased attribute/transformation enumeration depth is reduced encoding channel coverage, as shown in \autoref{tab:constraints}.

\vspace{-1mm}
\paragraph{Comparing Algorithms in Terms of Ranking Trade-Offs.} 
% We assess the overall complexity of the ranking step for each of the four algorithms in \autoref{fig:ranking-table}. 
Three of the four algorithms we compare utilize behavioral ranking models (Voyager, Dziban, DeepEye). These behavior-based heuristics are fast to apply to visualization candidates, but can take significant effort to derive on account of having to conduct user studies and/or field studies beforehand to collect the data~\cite{saket2018beyond}.
Even when the data is collected, significant manual effort may also be required to hand-tune the resulting models~\cite{Wongsuphasawat2015voyager,Wongsuphasawat2017voyager2,Lin2020dziban}. This issue of effort is also observed for machine-learning models, such as in the case of DeepEye, which required extensive data collection to train its machine-learning oracle. In the case of Voyager and Dziban, existing heuristics, algorithms, and user study data were used to develop the oracles, which can help reduce the burden of training and tuning new models.
%leading to more efficient training of these models, however, it might take a lot of effort to hand tune data in preparation.
Foresight's oracle requires no training since it relies on a pre-defined set of statistics. However, Foresight must calculate these statistics for all enumerated attribute combinations, making its execution more expensive. Foresight uses statistical sketches to reduce the processing time.

Once these algorithms are finally trained and tuned, a natural question is: which algorithm provides the best recommendations for a given visualization task? Though this question could be evaluated theoretically, existing approaches often use a somewhat reductive approach of approximating users' analytic performance through low-level perceptual heuristics (e.g., \cite{Luo2018deepeye,Lin2020dziban}). Perception is only one component of a user's analytic performance and is a poor approximation of user performance in higher-level visual analysis tasks, such as prediction or exploration~\cite{Battle2019characterizing}.
Instead, we argue for an empirical evaluation approach that is more task-sensitive.
%Although we could theoretically compare different existing recommendation algorithms in terms of enumeration and ranking, it is still hard to know which algorithm provides better recommendations. 
To compare the quality of generated recommendations, we provide a demonstration of using our framework to empirically evaluate different algorithms in the following sections.

\vspace{-1mm}
\section{Benchmarking Recommendation Algorithms}
\label{sec:user-study}
\vspace{-1mm}

% We conducted an exploratory user study to evaluate and compare the user performance of four new visualization recommendation algorithms (CompassQL+BFS, CompassQL+DFS, Dziban+BFS, Dziban+DFS) as a demonstration to show how our framework can guide the benchmarking of various recommendation algorithms.
%different combination of oracles (Dziban~\cite{Lin2020dziban} and CompassQL~\cite{Wongsuphasawat2016towards}) and traversal methods (breadth-first \& depth-first) 

% \leilani{mention algorithms are moving in the direction of hybrid recommendation~\cite{Wongsuphasawat2016towards}. We focus on hybrid techniques in our experiment (e.g., Voyager and Dziban).}

We show that our framework could compare a wide range of existing visualization recommendation algorithms \emph{theoretically} in
the previous section. 
Whereas in this section, we show how our framework could guide the \emph{empirical} comparison of various recommendation algorithms.

\vspace{-1mm}
\subsection{Algorithms for Standardized Evaluation}
\label{demonstration}
\vspace{-1mm}

% We already show that our framework could compare a wide range of existing visualization recommendation algorithms theoretically in \ref{sec:examples}. 
% We would like to further present how our framework could guide the empirical comparison of various recommendation algorithms.
% To evaluate and compare the user performance of various recommendation algorithms, a standardized interactive interface is needed. 
Existing recommendation algorithms either have no interface presented~\cite{Hu2019vizml, Lin2020dziban, Moritz2018formalizing} or the systems built on the top of them utilize different interface designs~\cite{Demiralp2017foresight, Key2012vizdeck, Luo2018deepeye, Vartak2015seedb, Wongsuphasawat2015voyager, Wongsuphasawat2017voyager2, Hu2018dive}, which makes it hard to conduct a standardized evaluation.
Moreover, various systems allow different kinds of user input, which brings even more difficulties to the evaluation and comparison.
For instance, the majority of systems allow selected data fields as input~\cite{Demiralp2017foresight, Key2012vizdeck, Luo2018deepeye, Wongsuphasawat2015voyager, Wongsuphasawat2017voyager2, Hu2018dive}, while some also allow inputting statistical features~\cite{Demiralp2017foresight}, or visualization types~\cite{Key2012vizdeck}.

% Thus, to standardize the evaluation and comparison of different recommendation algorithms, we do not evaluate existing systems as the way they are presented, but instead compare among new algorithms which apply the enumeration approach and oracle behind them.
Thus, to standardize the benchmark of different recommendation algorithms, we implement an interface to wrap around algorithms that are generated by applying the enumeration approach and oracle behind existing recommendations.
In this paper, by varying the traversal method and the oracle, we come up with four new visualization recommendation algorithms to evaluate.
The graph traversal method would be either BFS or DFS, and the oracles are CompassQL~\cite{Wongsuphasawat2016towards} and Dziban~\cite{Lin2020dziban}.
Both BFS and DFS are tree-oriented traversal methods.
While BFS enumerates with a maximum path length of one, DFS enumerates along the path until the current node or the space boundary is reached.
The CompassQL version that we use is the same as the one behind the Voyager~\cite{Wongsuphasawat2015voyager, Wongsuphasawat2017voyager2} systems, which ranks visualizations by effectiveness.
On the other hand, Dziban is built on the top of Draco~\cite{Moritz2018formalizing} and GraphScape~\cite{Kim2017graphscape}, which takes both effectiveness and perceptual distance into consideration while ranking visualizations.
%The reason that we pick these two oracles partly is that the source codes of other recommendation algorithms are not publicly accessible~\cite{Demiralp2017foresight, Key2012vizdeck, Luo2018deepeye, Hu2019vizml}.

\revised{We evaluate CompassQL~\cite{Wongsuphasawat2016towards} and Dziban~\cite{Lin2020dziban} based on the availability of their source code, whereas the code for many other} algorithms is not publicly accessible~\cite{Demiralp2017foresight, Key2012vizdeck, Luo2018deepeye, Hu2019vizml}.
Moreover, by adding the ranking strategy of GraphScape~\cite{Kim2017graphscape} to optimize the perceptual distance, Dziban~\cite{Lin2020dziban} claims to provide a considerable benefit \revised{over} Draco-CQL~\cite{Moritz2018formalizing}, which is a re-implementation version of CompassQL~\cite{Wongsuphasawat2016towards}.
\revised{We} benchmark these two ranking engines to see whether there exists a significant improvement \revised{in} user performance.
% As a result, we generate four new visualization recommendation algorithms, CompassQL+BFS, CompassQL+DFS, Dziban+BFS, and Dziban+DFS.
The visualization design space is the same for all algorithms, where each node contains visualizations with the same \revised{data attributes}, and each edge represents \revised{adding} or removing one data attribute.
We only consider visualizations with 3 data attributes or less, thus no attribute can be removed from a univariate chart, and no attribute can be added to a three-attribute chart.
Oracles would need to make other data variation decisions, like whether to add data transformations (binning or aggregating), as well as design variation decisions, like applying which visual encoding for each attribute.
% We then discuss the interface design and the user study design in \autoref{sec:user-study} and benchmark results in \autoref{sec:analysis}.

\vspace{-1mm}
\subsection{Interface Design}
\vspace{-1mm}

\begin{figure*}
\centering
 \includegraphics[width=0.85\textwidth]{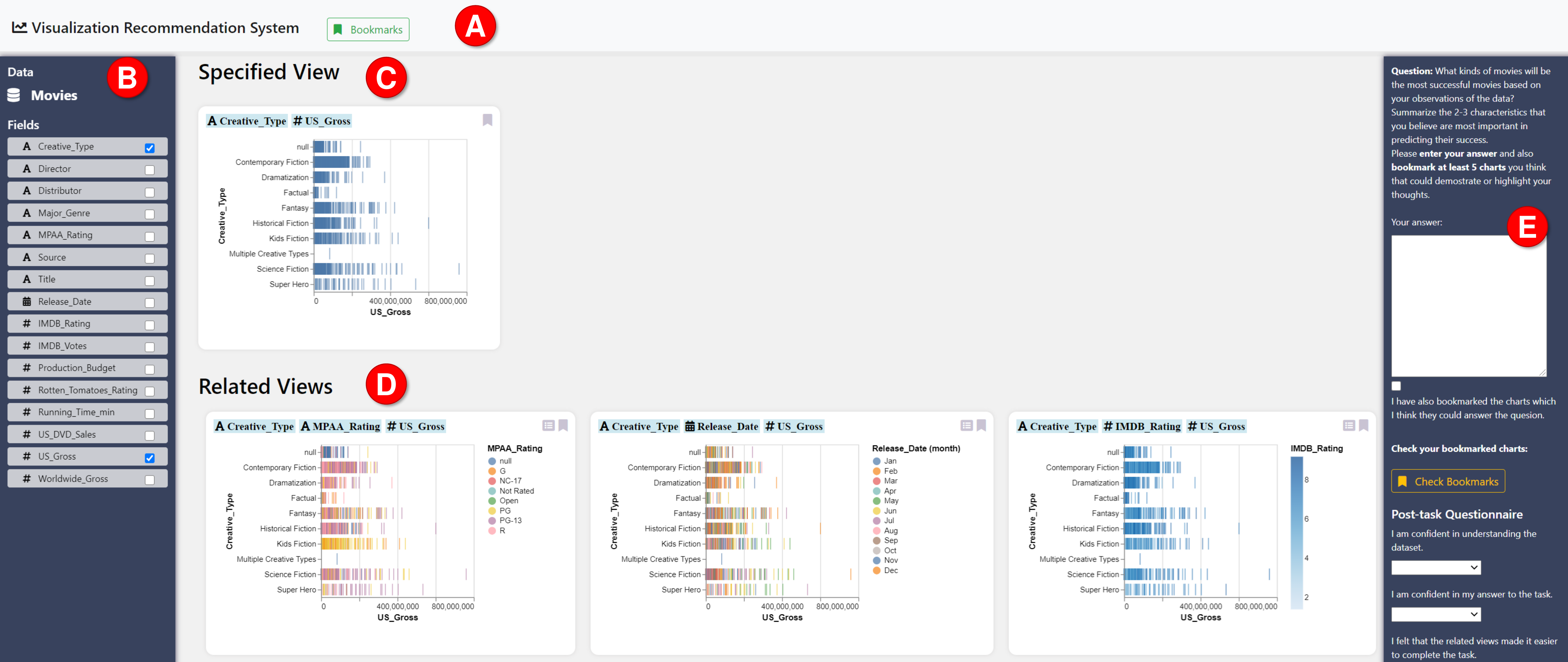}
 \vspace{-2mm}
 \caption{Interface for the user study. The \textbf{top panel (A)} provides the button to view the bookmark gallery. The \textbf{data panel (B)} contains the dataset name and data fields. Users can manually select which fields to be visualized. The visualization gallery contains the \textbf{specified view (C)} and \textbf{related views (D)}. The specified view displays the current specified chart while related views show recommended charts relevant to the specified chart. The \textbf{task panel (E)} contains the current task and also the post task questionnaire.}
 \label{fig:interface}
 \vspace{-5mm}
\end{figure*}
% \zehua{If we convert this figure to columnwidth, then we are on exactly 9 page.}
% We got the inspiration of the interface design from Voyager~\cite{Wongsuphasawat2015voyager, Wongsuphasawat2017voyager2}.

\autoref{fig:interface} shows the interface for evaluating the set of new visualization recommendation algorithms, which consists of a \textit{top panel}, a \textit{data panel} (left), a \textit{visualization gallery} (middle), and a \textit{task panel} (right).
Our interface design is inspired by the Voyager systems~\cite{Wongsuphasawat2015voyager, Wongsuphasawat2017voyager2}.
Since we focus on evaluating the recommendation quality of each algorithm, limited interactions are allowed in the interface, such as selecting attributes, bookmarking or specifying a chart, and also hovering over a chart to check the value of a particular data point. 
% We also provide the source code of our interface design in the OSF repository: {\small \url{https://osf.io/txqsu/?view_only=17a4bbf7239547f18305584c842c5a3c}}.
\revised{We share the source code and a demonstration video of our interface in the OSF repository.}

% \noindent\jane{Consider adding some more rationale to this section. For example, you might emphasize that the interface itself is not necessarily a new contribution (the goal is not to design the perfect new interface) but instead is a playground for testing recommendations. I also think it would be helpful to include some rationale for why the interface is designed this way (for example, why can't the user choose the marktype?).\par}

\begin{comment}
\begin{figure}[tb]
\centering
 \includegraphics[width=1.0\columnwidth]{pictures/bookmarks.pdf}
 \caption{A bookmark gallery of visualizations saved by an analyst. \leilani{Is there a concrete insight/takeaway for this figure?}}
 \label{fig:bookmarks}
\end{figure}
\end{comment}

% \vspace{-1mm}
% \subsubsection{The Top Panel}
% \vspace{-1mm}

\vspace{-1mm}
\paragraph{The Top Panel (A).}
By clicking the button in the top panel, a bookmark gallery of visualizations saved by the user pops up.
% \autoref{fig:bookmarks} shows an example of the bookmark gallery created by an analyst in one of the tasks.
Participants are encouraged to bookmark charts that could answer the question during the user study.
% There is a button for users to view the bookmark gallery in the top panel.

% \vspace{-1mm}
% \subsubsection{The Data Panel}
% \vspace{-1mm}

\vspace{-1mm}
\paragraph{The Data Panel (B).}
It shows the name of the current dataset and presents a list of all data fields within the dataset.
The list is grouped by the data type and then ordered by the variable name alphabetically.
For each variable, it shows the data type icon, the variable name and then a checkbox representing whether the variable is included in the specified view.
Users can click on the checkbox to include or exclude an attribute from the specified view (C). The related views (D) will provide different recommendations based on the current specification. 

\begin{comment}
\begin{figure}[tb]
\centering
 \includegraphics[width=1.0\columnwidth]{pictures/default.pdf}
 \caption{The related views would show all charts with 1-attribute when no data field is selected. \leilani{cut}}
 \label{fig:default}
\end{figure}
\end{comment}

% \vspace{-1mm}
% \subsubsection{The Visualization Gallery}
% \vspace{-1mm}

\vspace{-1mm}
\paragraph{The Visualization Gallery (C \& D).}
It consists of two views: the specified view~(C) and the related views~(D). 
% By default, right after the page loaded or no variable is selected, no chart would be shown in the specified view, while charts visualizing each data field would be shown in the related views for users to explore (\autoref{fig:default}).
Each chart contains a label on the top-left corner showing which data attributes are visualized in the chart, and a \textit{bookmark} button (\faBookmarkO) on the top-right corner, which triggers whether the bookmark gallery includes or excludes the chart.
% The \textit{list} button (\faListAlt) only exists in the chart in the related views.

%The specified view (C) shows the chart which visualizes the currently selected variables. 
The specified view (C) is the best chart recommendation for the currently selected variables.
% \jane{Is this a fair rewording?}
%The specified view can also be set by selecting a chart from the related views (D) using the corresponding \mbox{\textit{list} button (\faListAlt).}
%The specified view could be changed not only by changing the attribute selection in the data panel, but also by specifying a chart via clicking on the \textit{list} button (\faListAlt) in the related views.
The related views show recommended charts by the current recommendation algorithm based on the specified view.
% \jane{and traversal method}.
When no data field is selected, the related views show univariate visualizations.
By default, the related views display the top five recommended charts based on the specified view, if users want to explore more, they can click on the \textit{Load More} button (\faSpinner \, Load More) to view additional recommendations.
The \textit{list} button (\faListAlt) on each chart in the related views allows users to update the specified view and display new recommendations from this starting point ($n_0$).
%The \textit{list} button (\faListAlt) on each chart in the related views is for users to check the recommended visualizations generated based on the corresponding chart.
% The \textit{list} button (\faListAlt) is also shown on the charts in the related views.
% If users are interested in a chart and would like to see the recommended charts generated from it, they can click on the \textit{list} button (\faListAlt) to make it as the specified view.
% The selected variables, and the related views would be changed accordingly.

% \vspace{-1mm}
% \subsubsection{The Task Panel}
% \vspace{-1mm}

\vspace{-1mm}
\paragraph{The Task Panel (E).}
It consists of (1) the current task description, (2) an input area for users to answer the question, (3) a checkbox for users to self-check if they bookmarked charts that could help answer the question, (4) a button to revisit the bookmark gallery, (5) the post-task questionnaire, and (6) a submit button to navigate to the next step.
When participants click on the submit button, the answer of the task, the specifications of the bookmarked charts, the response of the post-task questionnaire, and also the interaction log will be sent to the server.

% \noindent\jane{I think I said this in a previous comment, but it would be great if this system is open source. If it is, include a placeholder URL in the submission and consider adding some brief additional information about how the evaluation results are saved or how questions are loaded into the system.\par}

\vspace{-1mm}
\subsection{Study Design}
\vspace{-1mm}

The study followed a 4 (recommendation algorithms) $\times$ 2 (dataset) mixed design, thus in total there are 8 designs.
We utilized a between-subjects study design;
each participant only conducted one analysis session, with a random combination of recommendation algorithm and dataset.
All participants completed the study remotely.
%The presentation order of different recommender design across subjects is random.

% \vspace{-1mm}
% \subsubsection{Visualization Tools}
% \vspace{-1mm}

\vspace{-1mm}
\paragraph{Visualization Tools.}
The \revised{interface (\autoref{fig:interface}) was the same for every participant, but the recommendation algorithm was varied to generate} different visualizations in the specified and related views.

% \vspace{-1mm}
% \subsubsection{Datasets}
% \vspace{-1mm}

\vspace{-1mm}
\paragraph{Datasets.}
We utilized two Voyager~\cite{Wongsuphasawat2015voyager} datasets for the evaluation: \emph{movies} and \emph{birdstrikes}.
The \emph{movies}\footnote{\tiny\url{https://github.com/vega/vega-datasets/blob/master/data/movies.json}} dataset contains 3,201 records and 15 attributes (7 nominal, 1 temporal, 8 quantitative). 
The \emph{birdstrikes}\footnote{\tiny\url{https://github.com/vega/vega-datasets/blob/master/data/birdstrikes.csv}} dataset is a redacted version of FAA wildlife airplane strike records with 10,000 records and 14 attributes (9 nominal, 1 temporal, 4 quantitative). 

% \noindent\jane{I tried to make the footnote urls smaller, but for some reason \textbackslash small and \textbackslash tiny work, but not  \textbackslash footnotesize or \textbackslash scriptsize, which are probably ideal. Maybe removing the indentation would be good too, but couldn't get that to work easily. \par}

% \vspace{-1mm}
% \subsubsection{Participants}
% \vspace{-1mm}

\vspace{-1mm}
\paragraph{Participants.}
% Since there are 8 different conditions, we need to have at least 8 participants for each condition to get meaningful results from the analysis.
We recruited nine subjects for each condition, for a total of 72 participants (23 female, 49 male), all of whom successfully completed the study and were included in our \revised{analysis}.
All participants claimed to have proficient computer skills and prior experience using at least one of the following or similar tools/programming languages: Excel, Tableau, Voyager, Python/matplotlib, R/ggplot, D3.
We recruited participants from both academia and industry.
Of the 72 participants, 40 were students while 32 were professional participants from the industry.
% In total, we have 40 student participants from the academic side and 32 professional participants from the industry. 
%Each participant completes 1 out of 8 designs. For 64 participants, we have 8 data points for each design.
%The learning curve will be shortened since prior data analysis experience is required. 
% The user study was between-subject, thus each participant completed one of the eight versions.
% Each study session lasted around 60 minutes.
We compensated participants with a \$10 Amazon gift certificate.

% \noindent\jane{How many participants were actually recruited? Section 5 states that there were 72 valid results and 12 participants omitted.\par}

% \vspace{-1mm}
% \subsubsection{Study Protocol}
% \vspace{-1mm}

\vspace{-1mm}
\paragraph{Study Protocol.}
Each participant completed a 60-minute online session, consisting of:
(1) study overview and consent form; 
(2) a demographic survey;
(3) 10-min tutorial and demo with a dataset distinct from those used for the actual analysis sections;
(4) 40-minute analysis block with one study design;
and (5) the exit-survey.
% Users are asked to think aloud, and verbalized summarize their findings of each task.
During the study session, participants were asked to complete four analysis tasks, two focused and two open-ended (see \autoref{tab:tasks}).
% They were asked to use the bookmark function to save charts that could provide the answers for the task.
After each task, participants were asked to reflect on their experience using the recommendation tool to complete the task in a short post-task questionnaire with a symmetric 5-point scale, from strongly disagree (-2) to strongly agree (+2): 

\begin{itemize}[noitemsep,topsep=1pt]
    \item \textbf{Confidence in Understanding Data:} I am confident in understanding the dataset.
    \item \textbf{Confidence in Answer:} I am confident in my answer to the task.
    \item \textbf{Efficiency:} \revised{The} \textit{related views} made it easier to explore the \revised{data}.
    \item \textbf{Ease of Use:} \revised{The} \textit{related views} were easy to understand.
    \item \textbf{Utility:} \revised{The} \textit{related views} were useful for completing the task.
    \item \textbf{Overall:} I would use this \revised{tool} for similar tasks in the future.
\end{itemize}
\revised{After completing all four tasks, participants completed} a survey to \revised{evaluate} the recommender. The following questions were \revised{asked}:
\begin{itemize}[noitemsep,topsep=1pt]
    \item What are the advantages and disadvantages of the tool?
    \item Do you have any other comments on the recommendation system?
\end{itemize}

% \jane{out loud or written text? ``interviewed'' implies out loud, whereas I usually think of ``exit-survey'' as something participants fill out.} 

% In the post-task questionnaire~\cite{Battle2020database}, we asked participants to rate the following six experience with a symmetric 5-point scale, from strongly disagree (-2) to strongly agree (+2): 
% We also asked the following two questions in the exit-study survey to gain more comments of the tool from participants: 

% \vspace{-1mm}
% \subsubsection{Tasks}
% \vspace{-1mm}

\begin{table*}
\caption{List of all task prompts used in the study for the Movies dataset.}
\vspace{-2mm}
\centering
\begin{tabular}{l|l}
\toprule
\small \textbf{Task ID} & \small \textbf{Task Prompt} \\
\midrule
\small \textbf{T1 (Find Extremum)} & \small Which creative type has the maximum number of movies based on Book/Short Story (Source)?\\
\rowcolor[HTML]{ededed} 
\small \textbf{T2 (Retrieve Value)} & \small Among Disney Ride (Source) movies, what is the running time (mins) of the highest Worldwide grossing movie? \\
\small \textbf{T3 (Prediction)} & \small \makecell[l]{What kinds of movies will be the most successful movies based on your observations of the data? Summarize the 2-3 characteristics that\\ you believe are most important in predicting their success.}\\
\rowcolor[HTML]{ededed} 
\small \textbf{T4 (Exploration)} & \small \makecell[l]{Feel free to explore any and all aspects of the data for up to [15 mins]. Use the bookmark features to save any interesting patterns, trends\\ or other insights worth sharing with colleagues.} \\
\bottomrule
\end{tabular}
\label{tab:tasks}
\vspace{-2mm}
\end{table*}

\vspace{-1mm}
\paragraph{Tasks.}
We designed four visual analytics tasks (see \autoref{tab:tasks}) for each dataset based on prior studies of data analysis~\cite{Battle2020database, Battle2019characterizing, Wongsuphasawat2015voyager, Wongsuphasawat2017voyager2}.
These four tasks cover all three analysis task classes discussed by Battle et al.~\cite{Battle2020database}: quantitative, qualitative, and exploratory.
T1 and T2 are focused tasks; T1 involves two data attributes, while T2 involves three data attributes.
T1 asks participants to find the extremum, which is a qualitative task, while T2 asks participants to retrieve a specific value from a subset of the data, which is a quantitative task.
Both T3 and T4 are exploratory tasks.
T3 provides a particular direction for the data exploration, while T4 asks participants to freely explore the dataset.

\vspace{-1mm}
\paragraph{Collected Data.}
Since the user study was conducted remotely, for each task, we collected participants' (1) answers, (2) bookmarked charts, (3) interaction logs, and (4) responses of post-task questionnaires.
We also obtained comments from the exit-study survey.
% We gathered all results to analyze and compare the performance of each visualization recommendation algorithm.

% \vspace{-1mm}
% \subsubsection{Pre-registration}
% \vspace{-1mm}

\vspace{-1mm}
\paragraph{Pre-registration.}
We pre-registered\revised{~\cite{Lakens2019value, Wicherts2016degrees, De2014meaning}} the conditions, measurements, analysis (using Bayesian regression models to test if there is a significant difference in the stated measurements), and data collection criteria on the website AsPredicted\footnote{\tiny\url{https://aspredicted.org/blind.php?x=2vi5tq}} before collecting any data.

\vspace{-1mm}
\section{Benchmark Results}
\label{sec:analysis}
\vspace{-1mm}

We obtained 72 valid study results which passed the exclusion criteria in our pre-registration.
% We omitted 12 participants who completed the study but did not pass the pre-registered criteria.
We also had a pilot study with five \revised{participants}, where we derived the informative priors for our quantitative analysis.

% As mentioned in the study protocol, during the study section, participants completed 4 analysis tasks, 2 focused and 2 open-ended oriented ones.
% We collected participants' answers, bookmarked charts, interaction logs, and responses from post-task questionnaires of each task, and comments from the exit-study survey.

We now present the analysis of study results, focusing on the accuracy and completion time of focused-tasks, user interaction activities during open-ended tasks, post-task questionnaire responses, and qualitative feedback.
For quantitative analyses, we adopted Bayesian models to estimate the 95\% credible interval (CI) for each parameter.
Since the data type of our collected data varies, we had to apply various Bayesian regression models.
We used the logistic regression for analyzing the accuracy, the linear regression for the completion time and interaction logs, and the ordinal regression for post-task questionnaire responses.
We chose Bayesian models because they allow us to draw more reasonable conclusions about the true values of our parameters from small-n studies than the null hypothesis significant testing (NHST).
The Bayesian 95\% credible interval represents the interval that we are 95\% sure contains the true value, which is different from the NHST confidence interval.
\revised{On the other hand, in terms of estimating differences, like the differences between design \textit{A} and \textit{B} (i.e. A-B), if the Bayesian 95\% credible interval is greater than 0 and not overlapping with 0, it means that we are 95\% sure that design \textit{A} performed better than \textit{B}.}
We provide our experiment code, data collected for both experiments, and analysis scripts as supplemental materials in the OSF repository.
% \footnote{\tiny\url{https://osf.io/txqsu/?view_only=17a4bbf7239547f18305584c842c5a3c}}.

\vspace{-1mm}
\subsection{Focused Tasks}
\vspace{-1mm}

We use accuracy and completion time as the two metrics to evaluate and compare the empirical utility of the four recommendation algorithms in supporting focus-oriented analysis.

\vspace{-1mm}
\subsubsection{Accuracy}
\vspace{-1mm}

To analyze task accuracy, we trained a Bayesian logistic regression model for the two focused tasks to model the probability of a correct answer given an oracle and graph traversal combination.

\begin{figure}[tb]
\centering
 \includegraphics[width=1.0\columnwidth]{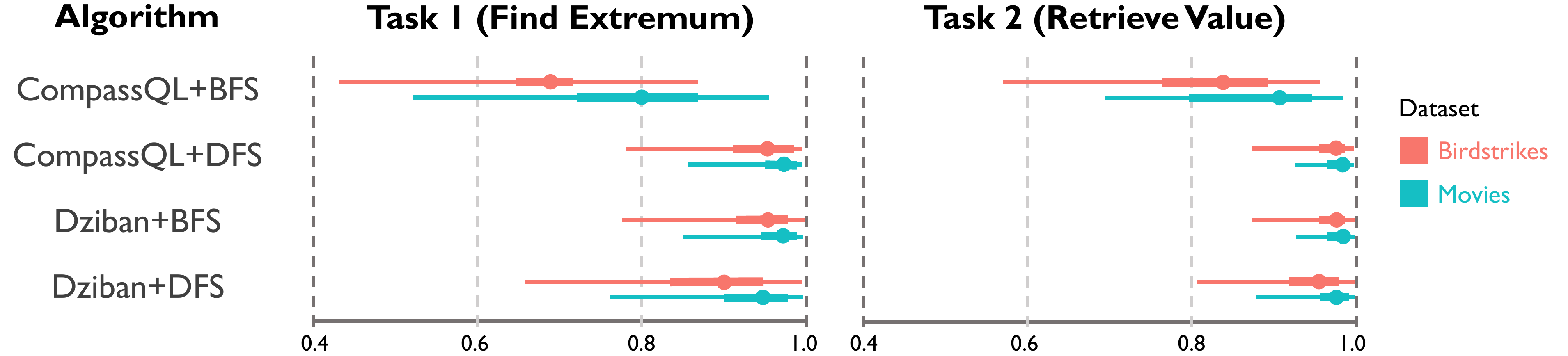}
 \vspace{-5mm}
 \caption{The predicted accuracy of focused tasks for all \revised{recommendation algorithms}. We show posterior distributions, 50\% and 95\% CIs of expected titer thresholds for both Movies and Birdstrikes dataset.}
 \label{fig:accuracy}
 \vspace{-2mm}
\end{figure}

% \autoref{fig:accuracy} shows weak evidence that CompassQL+DFS and Dziban+BFS have higher accuracy than Dziban+DFS, while CompassQL+BFS seems to have the lowest accuracy rate.
\revised{It shows in \autoref{fig:accuracy}} that CompassQL+DFS and Dziban+BFS \revised{had} higher accuracy than Dziban+DFS, while CompassQL+BFS \revised{seemed} to have the lowest accuracy rate.
However, since all of the 95\% CIs overlap, we cannot make a formal conclusion about which algorithm \revised{performed} significantly better in the accuracy of focused tasks. 

\vspace{-1mm}
\subsubsection{Completion Time}
\vspace{-1mm}

% \jane{What if you made the text for movies and birdstrikes the same color as the legend?}

\begin{figure}[tb]
\centering
 \includegraphics[width=1.0\columnwidth]{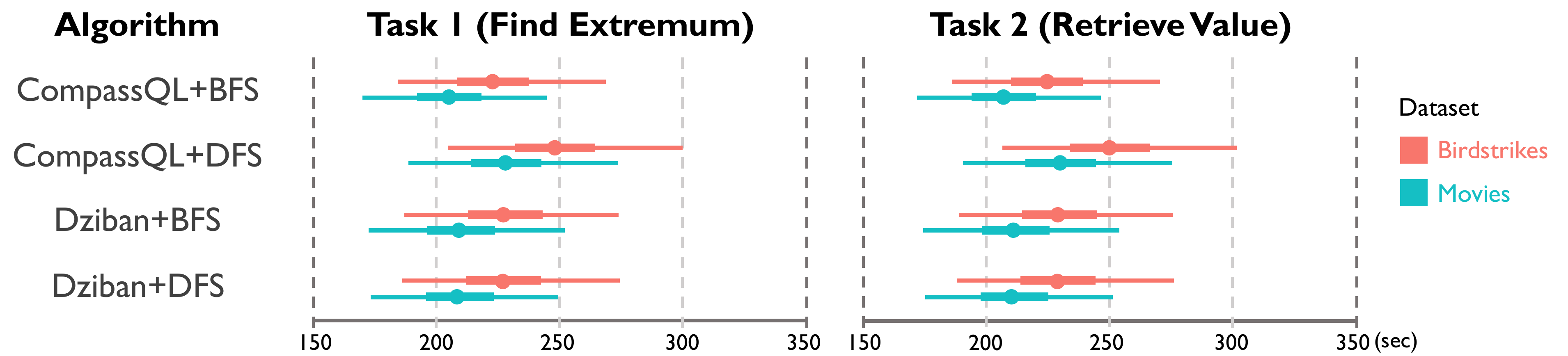}
 \vspace{-5mm}
 \caption{The completion time of focused tasks for all \revised{recommendation algorithms}. We show posterior distributions, 50\% and 95\% CIs of expected titer thresholds for both the Movies and Birdstrikes dataset.}
 \label{fig:time}
 \vspace{-5mm}
\end{figure}

We derived a weakly informative prior on completion times in seconds from the pilot study: N($\mu=360.48, \sigma=224.40$).
% Our priors are as follows,
% find-extremum: N($\mu=190.34, \sigma=96.57$),
% retrieve-value: N($\mu=313.37, \sigma=88.50$).
As shown in \autoref{fig:time}, all 95\% CIs overlap with each other, thus we cannot conclude which recommendation algorithm \revised{had} a significant effect on the completion time of focused tasks.
However, it is interesting to see that while participants spent the most time with CompassQL+DFS, the accuracy with CompassQL+DFS was the highest.
This relationship could imply that the longer time that participants spent in the task led to a higher accuracy.
% The result does show some differences among tested algorithms, however, since all 95\% CIs are overlapping, we cannot conclude that the difference of recommendation algorithms has a significant impact on the completion time for focused tasks.
Although not \revised{significant}, it generally takes less time for participants to complete tasks with the Movies dataset than the Birdstrikes one. On the other hand, the accuracy with the Movies dataset is also slightly higher.
This finding is reasonable, since people are more familiar with Movies data than Birdstrikes data in real life.

In summary, since all 95\% CIs overlap in both the accuracy and the completion time analysis, we conclude for preciseness and decisiveness that the four new recommendation algorithms have no significant impact on the performance of participants in focused tasks.

\vspace{-1mm}
\subsection{Open-ended Tasks}
\vspace{-1mm}

To evaluate the utility of different algorithms for supporting open-ended tasks, we analyze the interaction logs from the user study. 
Since the user study was conducted remotely, we lack eye-tracking data to show which visualizations users were attending to.
Taking inspiration from Voyager~\cite{Wongsuphasawat2015voyager, Wongsuphasawat2017voyager2}, we analyze the number of unique variable sets shown on screen to assess which recommendation algorithm provides broader data exploration during the open-ended tasks.
Moreover, we extend the analysis to the number of unique visual designs.
Unlike the variable set which only considers the combination of data fields, the visual design takes data transformations and visual encodings into account.
Since each edge in the visualization design space only represents the attribute modification, and oracles need to make choices for data transformations and encoding channels, it would be interesting to see whether the oracle would provide different visualization designs from the same node while the reference node ($n_0$) is different.

% \noindent\jane{While I like this paragraph, I want to flag it as a point for discussion similar to my previous comments about how the design space was defined. It is interesting to me that the design space focuses only on variable sets, whereas this analysis considers variable sets and encodings. (Note: I have not reviewed Zehua's most recent changes to those sections yet, so this point may be moot.)\par}

% Similar to Voyager~\cite{Wongsuphasawat2015voyager, Wongsuphasawat2017voyager2}, to assess which recommendation algorithm provides broader data exploration during the open-ended tasks, we analyze the number of unique variable sets (ignoring data transformations and visual encodings) that user were exposed to.
% In addition, we also would like to analyze the number of unique visual designs (taking variables, data transformations, and visual encodings into account) that are available to users during the open-ended tasks.

\begin{figure}[tb]
\centering
 \includegraphics[width=1.0\columnwidth]{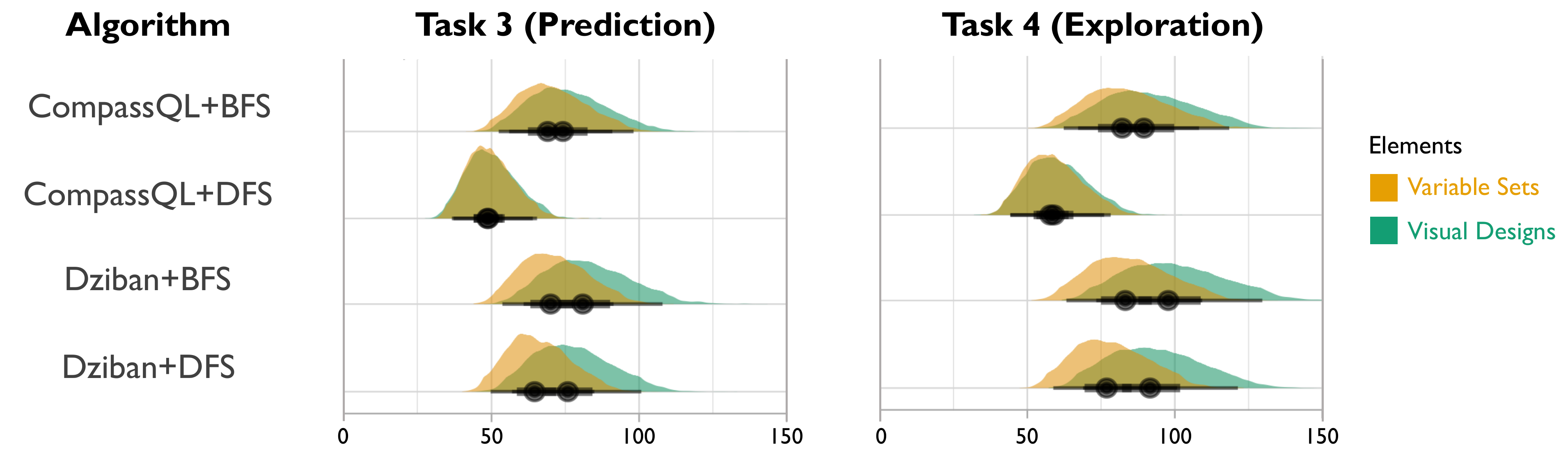}
 \vspace{-5mm}
 \caption{The number of exposed variable sets and visual designs of open-ended tasks among all \revised{recommendation algorithms}. We show posterior distributions, 50\% and 95\% CIs of expected titer thresholds.}
 \label{fig:exposed}
 \vspace{-2mm}
\end{figure}
% \jane{Minor: I propose changing the color for these charts so that the red/blue is only the dataset and other colors are used for the elements.}

\begin{figure}
\begin{subfigure}[b]{1.0\columnwidth}
    \centering
    \includegraphics[width=\textwidth]{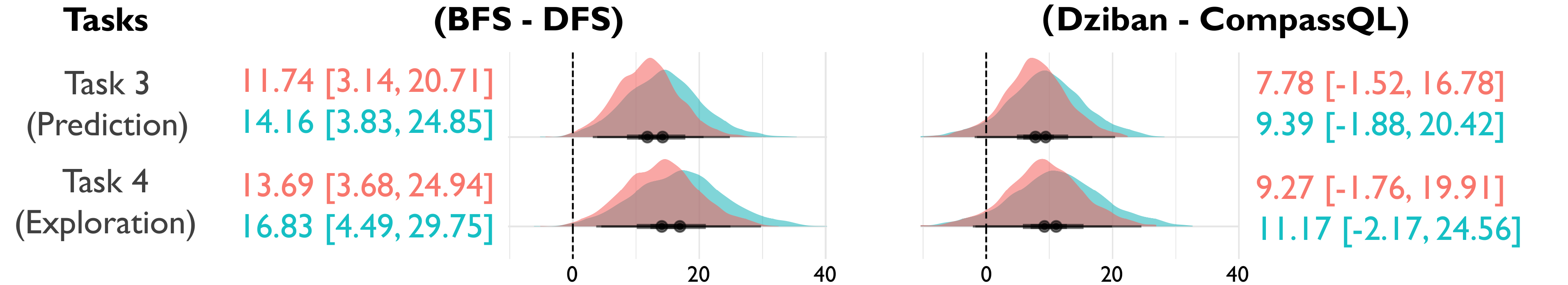}
    \vspace{-5mm}
    \caption{Difference in Average Numbers of Exposed Variable Sets}
    \label{fig:exposed-var-diff}
  \end{subfigure}
 \begin{subfigure}[b]{1.0\columnwidth}
    \centering
    \includegraphics[width=\textwidth]{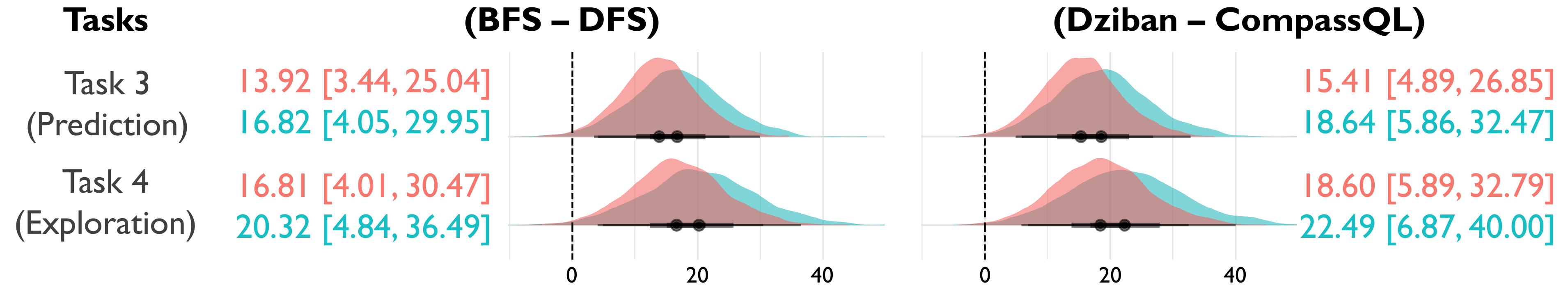}
    \vspace{-5mm}
    \caption{Difference in Average Numbers of Exposed Visual Designs}
    \label{fig:exposed-vis-diff}
  \end{subfigure}
  \vspace{-5mm}
 \caption{The differences in average numbers of exposed variable sets and visual designs. We show posterior distributions, 50\% and 95\% CIs of expected titer thresholds for both \crule[bird]{0.25cm}{0.25cm} Birdstrikes and \crule[movie]{0.25cm}{0.25cm} Movies dataset. }
 \label{fig:exposed-diff}
 \vspace{-5mm}
\end{figure}
% \jane{I like the changes to these figures, but what if we drop the legend from part (a) and also drop the caption text for (a) and (b) to save a little space? Instead, maybe add (a) and (b) to the figure image so they can still be referenced, perhaps ``(a) Tasks''}
% The left shows the difference between two traversal methods while the right shows the difference between two oracles.

\vspace{-1mm}
\subsubsection{Exposed Variable Sets \& Visual Designs}
\vspace{-1mm}

% \autoref{fig:exposed} shows us some evidence that BFS generally exposed more unique variable sets and visual designs than DFS, while Dziban also exposed more elements than CompassQL in the open-ended tasks.
\revised{\autoref{fig:exposed} shows that CompassQL+BFS, Dziban+BFS, and Dziban+DFS exposed more unique variable sets and visual designs than CompassQL+DFS in the open-ended tasks.}
On the other hand, we also see that Dziban exposed more numbers of visual designs than variable sets, so did BFS, which means Dziban and BFS recommended more design variants with the same variable sets, while CompassQL+DFS seemed to only recommend roughly one visual design for each variable set.
We also find that participants 
\revised{were}
% are
exposed to slightly more unique variable sets and visual designs in the exploration task than in the prediction task, which is reasonable since the exploration task encourages participants to explore the dataset freely while the prediction task restrains a direction for the data exploration.

To check the significance, we also run a Bayesian linear regression model on the exposure difference between BFS and DFS, as well as the difference between Dziban and CompassQL, as shown in \autoref{fig:exposed-diff}.

From \autoref{fig:exposed-var-diff} we can see that there is a significant difference in the average number of exposed variable sets between the two traversal methods, BFS and DFS.
In the prediction task, BFS exposed significantly more variable sets with both the Birdstrikes dataset ($b=11.746$) and the Movies dataset ($b=14.163$).
% \jane{focused?}
% On the other hand, in the exploration task, BFS exposed ($b=16.839$) more variable sets with the Movies dataset and ($b=13.965$) more with the Birdstrikes dataset.
We also find a similar pattern of the exposure difference in the exploration task that BFS exposed significantly more variable sets than DFS.
% between two traversal methods.
However, since the 95\% CIs overlap with the auxiliary line at 0, we cannot conclude that Dziban exposed significantly more unique variable sets than CompassQL.

% \noindent\jane{Consider rephrasing some of this paragraph using a form like ``BFS exposed more variable sets than DFS with both the Movies dataset ($b=14.163$) and the Birdstrikes dataset ($b=11.746$).''}

On the other hand, in terms of the number of visual designs, we find a significant difference between both traversal methods and between both oracles (\autoref{fig:exposed-vis-diff}).
BFS exposed significantly more visual designs with both the Birdstrikes dataset ($b=13.918$) and the Movies dataset ($b=16.822$) in the prediction task, while in the exploration task, BFS exposed ($b=16.813$) more with the Birdstrikes dataset and ($b=20.316$) more visualizations with the Movies dataset.
A similar pattern of the exposure difference is also found between the two oracles.
Dziban exposed significantly more visual designs than CompassQL with both the Movies and the Birdstrikes dataset in both tasks.
It is interesting to see that although Dziban did not expose significantly more unique variable sets, it exposed significantly more unique visual designs than CompassQL, which means Dziban tends to recommend more design variants than data variants (as shown in \autoref{fig:exposed}).
% This result is also verified in \autoref{fig:exposed}, which shows that Dziban exposed more unique visual designs than unique variable sets.

\vspace{-1mm}
\subsubsection{Interacted Variable Sets \& Visual Designs}
\vspace{-1mm}

\begin{figure}[tb]
\centering
 \includegraphics[width=1.0\columnwidth]{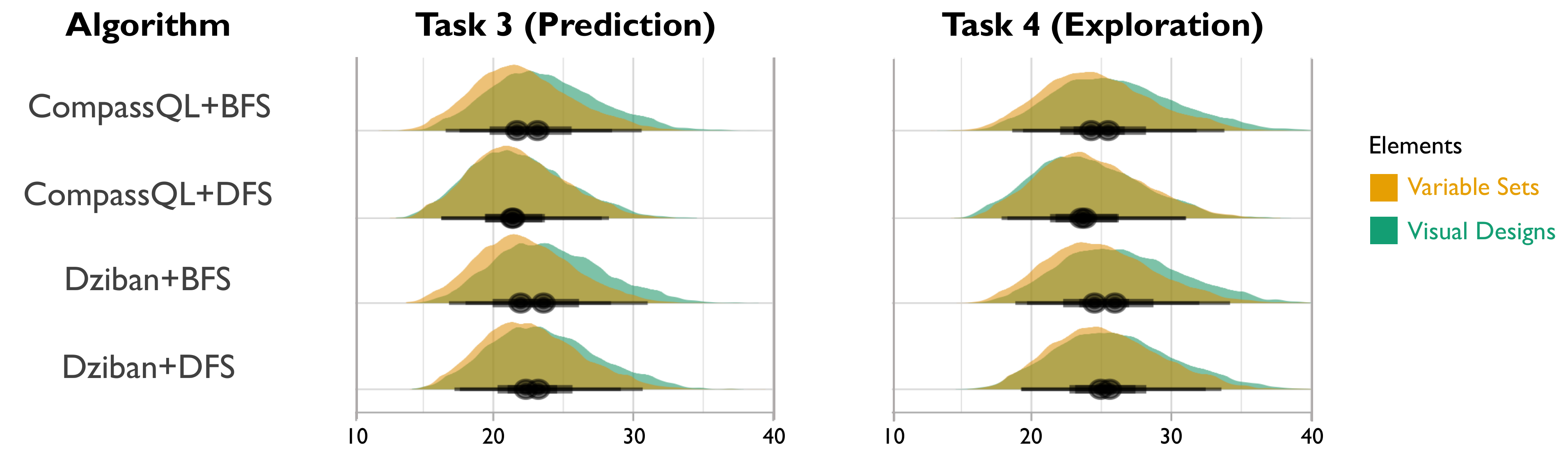}
 \vspace{-5mm}
 \caption{The number of interacted variable sets and visual designs of open-ended tasks among all \revised{recommendation algorithms}. We show posterior distributions, 50\% and 95\% CIs of expected titer thresholds.}
 \vspace{-2mm}
 \label{fig:interacted}
\end{figure}

We also analyze the number of unique variable sets and visual designs that participants interacted with during the open-ended tasks.
We include interactions like specifying (\faListAlt), bookmarking (\faBookmarkO), and mouse-hovering for more than half a second. 
% \jane{Did the charts have any basic interactions, like tooltips?}
From \autoref{fig:interacted} we do not see much difference in the number of interacted variable sets and visual designs among different recommendation algorithms.
It seems that participants interacted with more visual designs than variable sets with BFS, which means that BFS provides more interesting design variants that participants would like to interact with.
On the other hand, the number of unique variable sets and visual designs are about the same with DFS.
In other words, DFS did not expose as many interesting design variants as BFS (\autoref{fig:exposed}).

\vspace{-1mm}
\subsection{Post-task Questionnaires}
\vspace{-1mm}

We used the Bayesian ordinal regression model to analyze the user responses from the post-task questionnaires.
Since we used a symmetric 5-point scale (-2 strongly disagree, +2 strongly agree) in the post-task questionnaire, our prior on user score of (range $[-2, 2]$) is expressed as a normal distribution $N(0,1)$.

% \leilani{We need to collapse these so we have aggregated results by dfs/bfs, and results for task. This is too many rows otherwise.} 

\vspace{-1mm}
\subsubsection{Confidence Rating}
\vspace{-1mm}

\begin{figure}
 \includegraphics[width=1.0\columnwidth]{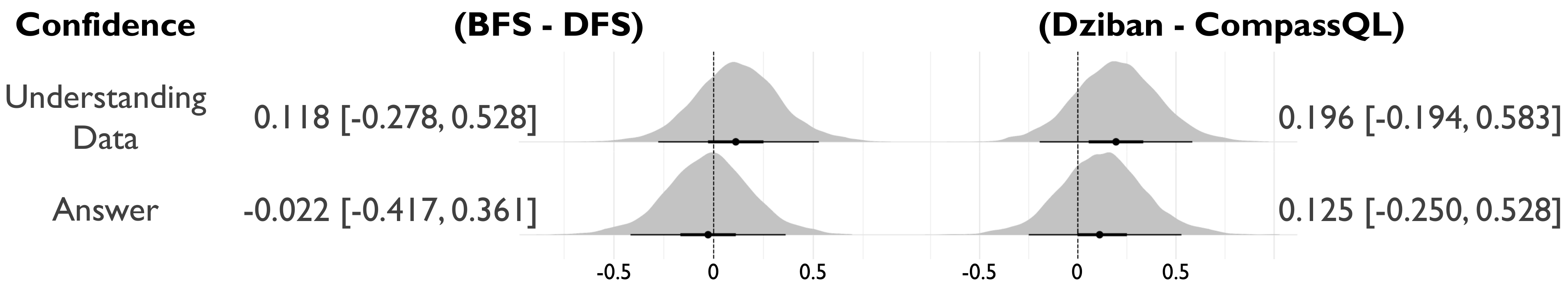}
 \vspace{-5mm}
  \caption{The differences in user confidence rating. We show posterior distributions, 50\% and 95\% CIs of expected titer thresholds.}
 \label{fig:confidence-diff}
 \vspace{-5mm}
\end{figure}

In the post-task questionnaire, we asked participants to rate their confidence in understanding data, and also in their answers.
% As shown in \autoref{fig:confidence-diff}, all of the 95\% CIs overlap with the auxiliary line at 0, we cannot conclude that the user confidence rating differs significantly among different search methods or oracles, but we still can find weak evidence from \autoref{fig:confidence-diff} supporting that Dziban performed a bit better than CompassQL towards user's confidence both in understanding data and in their answers.

\vspace{-1mm}
\paragraph{Confidence in Understanding data.}
    As shown in \autoref{fig:confidence-diff}, BFS performed slightly better than DFS on users' confidence in understanding data. \revised{Dziban} also had a higher confidence rating than CompassQL.

\vspace{-1mm}
\paragraph{Confidence in Answer.}
    On the other hand, BFS performed slightly worse than DFS on users' confidence in their answers.
    However, the Dziban oracle still had a slightly higher rating than CompassQL.
    
In summary, we don't see much difference in users' confidence ratings between the two traversal methods, BFS and DFS.
\revised{On the other hand, \autoref{fig:confidence-diff} shows that the Dziban oracle performed better than CompassQL, however, the outperformance was not significant.}
% On the other hand, \autoref{fig:confidence-diff} shows weak evidence (although not significant) that the Dziban oracle performed better than CompassQL with users' confidence in both understanding data and their answers.

\vspace{-1mm}
\subsubsection{Recommendation Algorithm Preference}
\vspace{-1mm}

\begin{figure}
 \includegraphics[width=1.0\columnwidth]{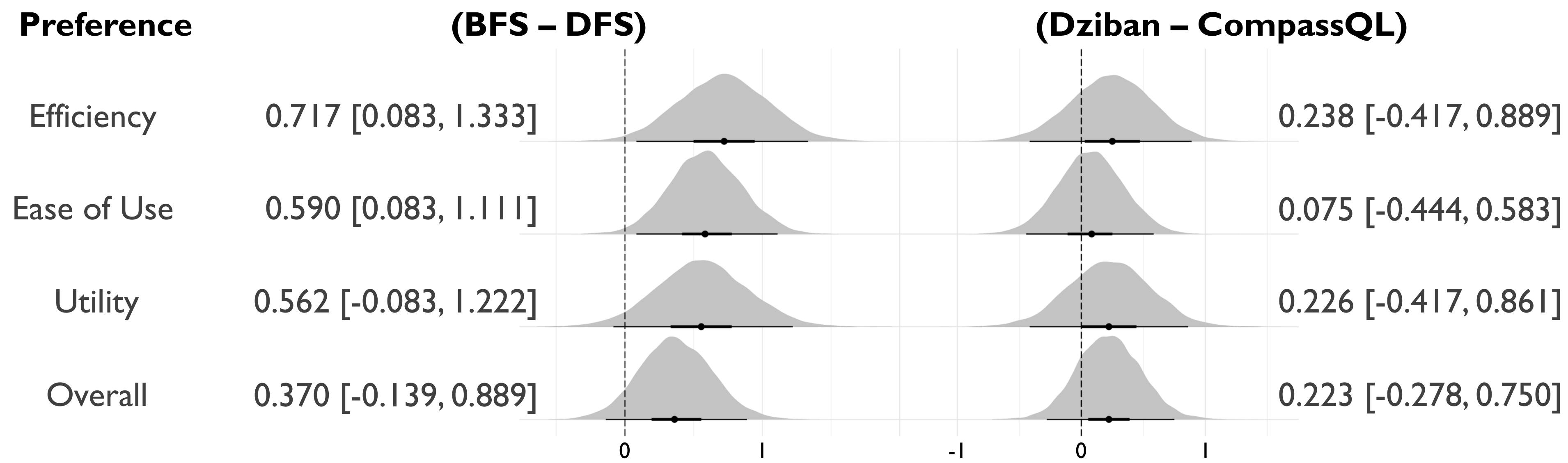}
 \vspace{-5mm}
  \caption{The differences in user preference rating. We show posterior distributions, 50\% and 95\% CIs of expected titer thresholds.}
 \label{fig:preference-diff}
 \vspace{-5mm}
\end{figure}

We also asked participants to rate the related views in different aspects: efficiency, ease of use, utility, and overall (\autoref{fig:preference-diff}). 
% As shown in \autoref{fig:preference-diff}, participants preferred BFS traversal method to DFS and Dziban oracle to CompassQL in all ratings, while some are significantly, some are not.

\vspace{-1mm}
\paragraph{Efficiency.}
    In terms of the efficiency rating, BFS received a significantly higher rating than DFS.
    However, although Dziban received a slightly better rating than CompassQL, we cannot conclude that a significant difference exists between the two oracles in the efficiency experience since the 95\% CIs overlap with the line at 0.

\vspace{-1mm}
\paragraph{Ease of Use.}
    Similar to the efficiency rating, BFS performed significantly better than DFS with respect to the ease of use.
    However, there is no evidence supporting that Dziban performed significantly better than CompassQL in terms of the ease of use.

\vspace{-1mm}
\paragraph{Utility.}
    Both 95\% CIs overlap with the auxiliary line at 0, thus we cannot conclude which traversal method or oracle is significantly better than the other one, although we can see that BFS and Dziban have slightly higher utility ratings than DFS and CompassQL respectively. 

\vspace{-1mm}
\paragraph{Overall.}
    Similar to the utility, since both 95\% CIs overlap with the line at 0, users did not significantly prefer one traversal method or oracle over the other one.
    
In summary, we see some significant differences in the user preference between the two traversal methods.
In particular, participants significantly preferred BFS in the efficiency and ease of use experience.
Although not significantly, we still can see that Dziban received a slightly higher rating in each experience compared to CompassQL.

\vspace{-1mm}
\subsection{Participant Feedback}
\vspace{-1mm}

\paragraph{DFS is not preferred for focused tasks.}
Participants dislike~DFS for focused tasks since the recommended charts could have more fields added compared to the current specified chart.
One participant found that \textit{``The recommendations were useful but most of the time distractive and too many for answering specific questions.''}
Another said that \textit{``When I checked one attribute, the recommendation charts always include three attributes. I would prefer if it was only two factors for the first two [focused] tasks.''}
However, when it comes to open-ended tasks, participants had a different point of view in terms of the DFS traversal method;
one participant mentioned that \textit{``This tool is good for exploring the data, especially it is the first time seeing (the data).''}

% \noindent\jane{Are the parts of the quotes in parentheses part of the original quote or added for clarity? I am more familiar square brackets [] for editing the clarity, e.g., ``for the first two [focused] tasks''.}

\vspace{-1mm}
\paragraph{Dziban is preferred as an oracle.}
Since Dziban takes the perceptual distance from the current chart into account, the behavior makes more sense to participants when they explore the related views.
One participant commented on CompassQL that \textit{``(I am) unsure if there is any logic on the recommended charts, sometimes they are completely useless and just layer on another random metric or dimension''}.
Another participant also pointed out that \textit{``The recommendations (from CompassQL) were often ineffective and created out of unrelated fields''}.
When it came to the recommendations from Dziban, participants provided more positive feedback.
One said that \textit{``The tool helps explore datasets and provides useful recommendations in terms of related measures and dimensions to enable getting useful insights.''}
Another participant also commented that \textit{``For the most part, the tool added fields that made sense to include in addition to the original choices.''}

\vspace{-1mm}
\paragraph{Both oracles need to be improved.}
Overall, we got positive feedback about the recommended charts, however, we also found some comments about the disadvantages of both oracles.
The most common issue for CompassQL is it recommends scatterplots a lot since it only emphasizes effectiveness and when it comes to three attributes, it picks \textsl{area} or \textsl{color} as the third encodings, which sometimes confuses participants.
One commented that \textit{``I didn't understand the shaded circle. I guess it could be there are various different values that are big and small.''}
In terms of color encoding, one participant commented that \textit{``(It seems to be) often picking categories to represent color where there were so many colors as to make them all meaningless''}, and another pointed out that \textit{``Colors did not seem related to essential data.''}
On the other hand, Dziban also considers the perceptual distance thus it tends to recommend charts that look similar to the original one but does not consider the effectiveness enough, like using text as a mark type in a scatterplot.
A participant commented on Dziban that \textit{``Don't recommend views where a text value would dominate the visualization.''}
% It would be an improvement that both oracles include more hand-tuned constraints.
% One of the potential improvements 
\revised{One way} to address these problems would be including more hand-tuned constraints, such as not using \revised{color} to visualize more than a certain number of categories, not using area encoding when the overlapping exists, and not using text as a mark type to visualize long content.

% \noindent\jane{I think it would be nice to have a final concluding sentence (or two) on this paragraph. In particular, what are some recommendations for how the recommendations could be improved and how would this fit into the framework (probably the ranking step)? How might you summarize the takeaways of this paragraph? \par}
\vspace{-1mm}
\section{Discussion \& Future Work}
\label{sec:discussion}
\vspace{-1mm}

In this paper, we presented an evaluation-focused framework that can describe many existing visualization recommendation algorithms, and showed how our framework could guide the theoretical and empirical comparison of such algorithms. We conclude this paper with a discussion of guidelines for new recommendation algorithms, key benchmarking takeaways, limitations, and opportunities for future work.
%Furthermore, we go on to describe the takeaways from the benchmarking, as well as the current limitations and opportunities for how future work can address those limitations.

\vspace{-1mm}
\subsection{The Framework As Guidelines}
\vspace{-1mm}
We now discuss how our framework could serve as a guideline not only for the future construction of recommendation algorithms but also for benchmarking a larger range of existing recommendation systems.

\vspace{-1mm}
\subsubsection{For Future Recommendation Algorithms}
\vspace{-1mm}

Our framework consists of three major components: (1) the visualization design space, (2) the traversal method, and (3) the oracle.
While constructing new recommendation algorithms, one should think about whether any of the components in the algorithm is new to the community.
For example, \textit{does my visualization design space contain more (meaningful) visual designs than other existing automated systems?} \textit{Is my algorithm using a new way to traverse the visualization space which could help the actual analysis?} \textit{Is there a new creative ranking strategy that has not been covered by the existing literature?}
On the other hand, thinking about different combinations of the three components is another creative opportunity for constructing new recommendation algorithms.
Among the new algorithms evaluated in this work, CompassQL+DFS, Dziban+BFS and Dziban+DFS have not previously been proposed to the community, although the ranking engines (CompassQL and Dziban) have been researched as key contributions in this space.

\vspace{-1mm}
\subsubsection{For Benchmarking Various Automated Systems }
\vspace{-1mm}

Although we did not benchmark existing automated systems since they leverage different interfaces and have limited code availability,
% \jane{fair rephrasing?}, 
%as the way they are developed \jane{What does this mean?},\zehua{like not using the same interface that they are using} 
our user study design still provides an at-a-glance overview of how our framework could be used to guide the evaluation and comparison of various automated systems.
Without a standardized interface design and style of user inputs, it is difficult to compare multiple recommendation algorithms.
By leveraging our framework, one could compartmentalize the three main components of the algorithm and test them within our standardized interface.
%One could fit various recommendation algorithms into our framework by extracting the three main components from the algorithm, and then implement them with the same interface design.
For instance, one of our proposed algorithms, CompassQL+BFS utilizes the same idea of the graph traversal method and the ranking engine behind the Voyager systems.
In such a way, our framework could not only evaluate the recommendation algorithm as a whole but also compare different components.
As shown in the previous section (\autoref{sec:analysis}), our results not only show which algorithm performed better but also which traversal method or oracle was preferred.

\vspace{-1mm}
\subsection{Takeaways from Benchmarking}
\vspace{-1mm}

% \jane{Remove this subheading (6.2.1) if we remove section 6.2.2}

% \subsubsection{Performance VS Preference}
% \vspace{-1mm}

From \autoref{sec:analysis}, we can see that there is actually no significant difference between recommendation algorithms in the participants' performance with focused tasks.
On the other hand, for open-ended tasks, we find that BFS exposed significantly more unique variable sets and visual designs than DFS, while Dziban exposed significantly more unique visual designs than CompassQL, but not variable sets.
However, when it comes to interacted variable sets and visual designs, we do not see any significant difference between BFS and DFS traversal methods and between Dziban and CompassQL oracles.
This finding raises an important point: significantly more exposure does not necessarily lead to significantly more interactions.
When designing a new visualization recommendation algorithm, exposing more data variants and design variants is a good trend.
However, if more exposure does not lead to more interactions, the resulting recommendations may lack the right level of ``interestingness'' for a worthwhile data exploration experience.

On the other hand, in terms of participants' preferences, we do find that participants significantly prefer BFS over DFS in the utility and ease of use ratings. 
Participants also prefer Dziban rather than CompassQL in all metrics (efficiency, utility, ease of use, and overall), although the rating difference is not significant.
Participants' post-study feedback also reveals their preference for Dziban as an oracle.
Since Dziban takes the perceptual distance into account, participants could better understand why such visualizations are recommended.

As we mentioned before, Dziban is an improved version of Draco-CQL (a re-implementation of CompassQL), which takes the perceptual distance into account.
However, we do not find a significant difference in the user performance between Dziban and CompassQL in focused-tasks, and the only significant improvement in open-ended tasks is that more visual designs are exposed (but not necessarily interacted with) 
% \jane{edited this sentence somewhat, is this fair (particularly the parenthetical)?}.
While the Dziban paper did present a comparison with Draco-CQL and GraphScape, it did not consider the user performance.
Based on their benchmark results, they claimed that Dziban provides a considerable benefit by suggesting charts that are effective, but also perceptually closed to the current one.
Nevertheless, without a framework to evaluate and compare the user performance between algorithms, we do not know whether the benefit would carry over into the actual analysis process.

From another perspective, we also find that users' preferences change with different analysis tasks, which implies that it is hard for a single algorithm to perform well across all tasks. 
When designing a new recommendation algorithm, one should think about which type of task to prioritize based on the expected goals of the intended users.
Alternatively, the recommendation system could switch to different algorithms depending on the particular task that users want to accomplish.

\vspace{-1.5mm}
\subsection{Limitations \& Future Work}
\vspace{-1mm}

Given the necessary level of visualization and analysis expertise for our participants, our recruitment protocol could not leverage standard crowdsourcing platforms, which limited the number of participants that we could feasibly recruit.
As a result, we limited our evaluation to two traversal methods and two existing ranking engines, CompassQL and Dziban.
However, it would be exciting to involve other promising traversal methods and oracles in future evaluations.

% \noindent\jane{This may be the perfect place to argue that our framework should make it possible for new systems to more easily compare to prior research by leveraging the same evaluation protocol, in which case we need to make sure that the full evaluation protocol is available for use by others. In other words, we can help support incremental evaluation of such recommendation algorithms.\par}

Given the current COVID-19 restrictions, the entire study was conducted remotely, which made it difficult to fairly perform a longer study session (like the 2-hour session in Voyager's evaluation~\cite{Wongsuphasawat2015voyager, Wongsuphasawat2017voyager2}).
% \jane{add citation here}
Therefore, we took a step back and chose the between-subjects study design, where each participant was exposed to only one recommendation algorithm.
However, the result would be more accurate and comparable if we could have conducted a within-subjects study.

As the study session length is limited, we could only pick a small number of visual analytics tasks to evaluate the user performance, while there exists a larger group of analysis activities in real-life practice~\cite{Kandel2012enterprise, Amar2005lowlevel}.
Moreover, our benchmark results imply that analysts prefer different algorithms for different analysis tasks.
Thus, one of the promising future work directions would be to include more analysis tasks into the benchmarking to better understand how different algorithms affect the user performance in various analysis tasks.
% separate tasks into low-level analysis tasks and high-level sensemaking ones and then benchmark the user performance in different ways.
% For example, recruiting participants on crowdsourcing platforms for low-level tasks while recruiting participants with higher-level expertise for complicated ones.

In this work, we focused on researching how different recommendation algorithms would affect the performance, behavior, and preference of participants, thus we only included limited interactions in our interface design.
However, from the post-study interviews, we find that participants would like to see the interface include more robust functionality, like filtering or supporting user-specified aggregations.
It would be interesting to see how the participant performance, particularly when interacting with charts, changes with those extra features, and whether such features would significantly affect the overall study results.
Since the source code for our empirical evaluation is publicly accessible, it would be easier to accomplish the aforesaid incremental evaluations.

\vspace{-1mm}

% We proposed three major axes for our framework, visualization space, traversal method and oracle. 
% During the process of constructing the new set of visualization recommendation algorithms, we find that there could be a couple of variants for the traversal methods, like we can split the DFS traversal method into DFS-path and DFS-leaf-node.
% It would be interesting to see that how we can vary the three major axes, or define more detailed sub-categories under each axe to describe more existing visualization recommendation algorithms and construct new algorithms.

% Among the feedback of the user study, we also received some comments on how to improve the user study in the future.
% Since we only want to research how different recommendation algorithms would affect the performance, behaviour, and preference of participants, thus we did not include other features into our interface design, like filtering.
% It would be also interesting to see how participant performance with those extra features.
% \input{content/conclusion}

\acknowledgments{
\vspace{-1mm}
The authors wish to thank the HCIL, the BAD Lab, and our paper reviewers for their thoughtful feedback. This work was supported in part by NSF award IIS-1850115 and an Adobe Research Award.}

\bibliographystyle{abbrv-doi}

\bibliography{reference}
\end{document}